\documentclass[conference]{IEEEtran}
\IEEEoverridecommandlockouts
\usepackage{libertine}
\usepackage{enumitem}
\usepackage{algpseudocode}
\usepackage[ruled,linesnumbered]{algorithm2e}
\usepackage{xcolor}
\usepackage{graphicx} 
\usepackage{epstopdf}
\usepackage{subfigure}
\usepackage{amsmath,bm}
\usepackage{multirow}
\usepackage{url}
\usepackage{amssymb}
\usepackage{bbding}
\usepackage{marginnote}

\renewcommand{\Comment}{}

\newcommand{\Revision}{}
\newcommand{\addComment}{}
\newcommand{\CHENG}{}
\newcommand{\ICDERevision}{}
\newtheorem{problem}{Problem}

\newcommand{\eg}{\textit{e}.\textit{g}.}

\begin{document}

\title{Online Anomalous Subtrajectory Detection on Road Networks with Deep Reinforcement Learning}

\author{\IEEEauthorblockN{Qianru Zhang\IEEEauthorrefmark{1}\IEEEauthorrefmark{3} \IEEEcompsocitemizethanks{\IEEEcompsocthanksitem\IEEEauthorrefmark{1}Both authors contributed equally to this research.},
Zheng Wang\IEEEauthorrefmark{1}\IEEEauthorrefmark{4}, 
Cheng Long\IEEEauthorrefmark{2}\IEEEauthorrefmark{4} \IEEEcompsocitemizethanks{\IEEEcompsocthanksitem\IEEEauthorrefmark{2}Corresponding author.},
Chao Huang\IEEEauthorrefmark{3},
Siu-Ming Yiu\IEEEauthorrefmark{3},
Yiding Liu\IEEEauthorrefmark{5},\\
Gao Cong\IEEEauthorrefmark{4},
Jieming Shi\IEEEauthorrefmark{6}
}
\IEEEauthorblockA{\IEEEauthorrefmark{3}Department of Computer Science, The University of Hong Kong, Hong Kong SAR\\
\IEEEauthorrefmark{4}School of Computer Science and Engineering, Nanyang Technological University, Singapore,
\IEEEauthorrefmark{5}Baidu Inc, China\\
\IEEEauthorrefmark{6}Department of Computing, The Hong Kong Polytechnic University, Hong Kong SAR\\
\{qrzhang,chuang,smyiu\}@cs.hku.hk, zheng011@e.ntu.edu.sg, \{c.long,gaocong\}@ntu.edu.sg} liuyiding.tanh@gmail.com, jieming.shi@polyu.edu.hk}


\maketitle

\pagenumbering{arabic}
\setcounter{page}{1}

\begin{abstract}
Detecting anomalous {\Comment{trajectories}} has become an important task in many location-based applications. 
While many approaches have been proposed for this task, they suffer from various issues including (1) incapability of detecting anomalous subtrajectories, which are finer-grained anomalies in trajectory data, and/or (2) non-data driven, and/or (3) requirement of sufficient supervision labels which are costly to collect.
In this paper, we propose a novel reinforcement learning based solution called \texttt{RL4OASD}, which avoids all aforementioned issues of existing approaches. 
\texttt{RL4OASD} involves two networks, one responsible for learning features of road networks and trajectories and the other responsible for detecting anomalous subtrajectories based on the learned features, and the two networks can be trained iteratively without labeled data. 
%
Extensive experiments are conducted on two real datasets, and the results show that our solution can significantly outperform the state-of-the-art methods (with 20-30\% improvement) and is efficient for online detection (it takes less than 0.1ms to process each newly generated data point). \\\vspace{-0.1in}

\end{abstract}



\renewcommand{\thesection}{\Roman{section}}
\setcounter{section}{0}

\section{INTRODUCTION}
\label{sec:introduction}

With the advancement of mobile computing and geographical positioning techniques, such as GPS devices and smart phones, massive spatial trajectory data is being generated by various moving objects (\eg, people, vehicles) in spatial spaces. Such collected trajectory data records the mobility traces of moving objects at different timestamps, which reflects diverse mobility patterns of objects. Accurate spatial trajectory data analysis serves as the key technical component for a wide spectrum of spatial-temporal applications, including intelligent transportation~\cite{yuan2021effective}, location-based recommendation service~\cite{feng2020hme}, pandemic tracking~\cite{seemann2020tracking}, and crime prevention for public safety~\cite{2018deepcrime}.



\begin{figure}
  \centering
  \includegraphics[width=0.68\linewidth]{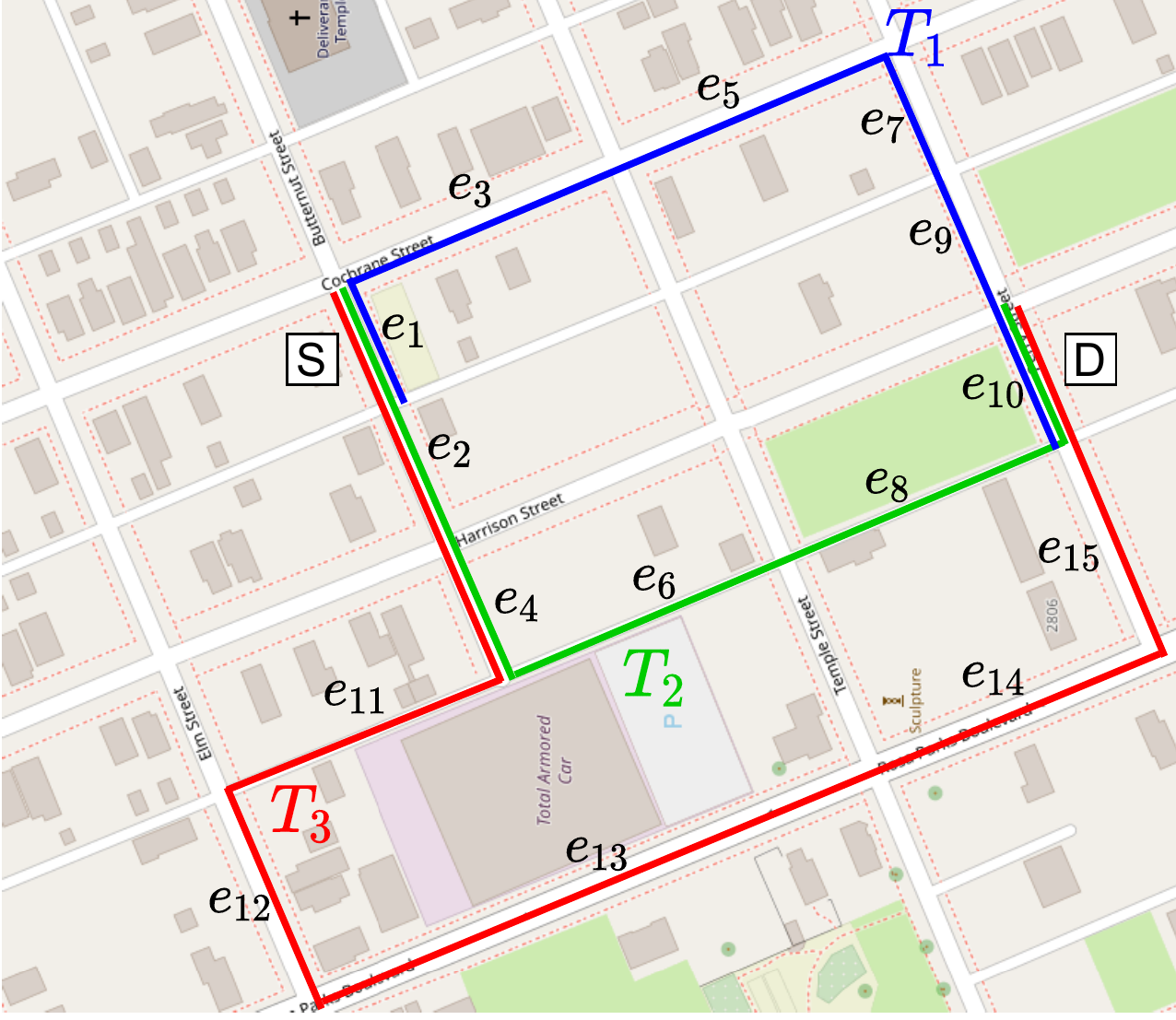}
  \caption{An example of anomalous trajectory. $T_1$, $T_2$ and $T_3$ are three trajectories from the same source $e_1$ to the same destination $e_{10}$. 
  }
  \label{fig:example}
  \vspace{-5mm}
\end{figure}

Among various trajectory mining applications, detecting anomalous trajectories plays a vital role in many practical scenarios. For example, the real-time vehicle trajectory anomaly detection is beneficial for better traffic management ~\cite{yuan2018hetero}. Additionally, studying the mobility behaviors of humans for discovering their anomalous trajectories is helpful for predicting event outliers (\eg, civil unrest and pandemic outbreaks)~\cite{ning2019spatio}. In such context, an outlier/anomaly refers to a trajectory which does not show the normal mobility patterns and deviates from the majority of the trajectories with the same source and destination~\cite{chen2011real, chen2013iboat} (termed as SD pair).
Consider the example in Figure~\ref{fig:example}. There are three trajectories from the source $S  (e_1)$ to the destination $D (e_{10})$. The trajectory $T_3$ (the red one) is considered as anomalous if the majority of the trajectories with the same SD pair $<S, D>$ follow either $T_1$ (the blue one) or $T_2$ (the green one). 

Recently, detecting anomalous trajectories has attracted much attention, and many efforts have been devoted to proposing various methods~(\cite{chen2013iboat, wu2017fast, zhang2020continuous, liu2020online,song2018anomalous,li2007roam}) for solving this problem. However, several key challenges have not been well addressed in current methods, which are summarized in Table~\ref{tab:summary} and elaborated as follows.


\begin{table}[]
\setlength{\tabcolsep}{4.5pt}
\centering
\caption{The existing methods and their corresponding issues ($\surd$ indicates the issue exists and $\times$ otherwise).}
\vspace*{-2mm}
\begin{tabular}{cccc}
\hline
 & \begin{tabular}[c]{@{}l@{}}Incapability of detecting \\ anomalous subtrajectories\end{tabular} & \begin{tabular}[c]{@{}l@{}}Non-data \\driven \end{tabular} & \begin{tabular}[c]{@{}l@{}}Requirement of sufficient  \\ supervision labels\end{tabular} \\ \hline
\cite{chen2013iboat} &$\times$  &$\surd$  &$\times$ \\
\cite{wu2017fast} &$\surd$  &$\times$  &$\times$ \\ 
\cite{zhang2020continuous} &$\surd$  &$\surd$  &$\times$ \\
\cite{liu2020online} &$\surd$  &$\times$  &$\times$ \\ 
\cite{song2018anomalous} &$\surd$  &$\times$  &$\surd$ \\ 
\cite{li2007roam} &$\surd$  &$\times$  &$\surd$ \\\hline
\end{tabular}
\vspace*{-4mm}
\label{tab:summary}
\end{table}

(1) \textbf{Incapability of detecting anomalous subtrajectories}. Most existing approaches {\Revision{~\cite{wu2017fast, zhang2020continuous, liu2020online, song2018anomalous,li2007roam}}} have only focused on identifying anomalous trajectories at coarse-grained levels, and detecting whether a trajectory \emph{as a whole} is anomalous or not.
However, the anomalous trajectory detection at the fine-grained level is more beneficial for better decision making in smart city applications, but unfortunately less explored in current methods. For instance, a ride-hailing company can immediately spot an abnormal driver when his/her trajectory starts to deviate from the normal route, which indicates an anomalous \emph{subtrajectory}.
Hence, in this work, we propose to detect fine-grained anomalous subtrajectories in a timely manner.



{\Revision{(2) \textbf{Non-data driven.}
Some existing approaches are based on pre-defined parameters and/or rules, and are not data driven~\cite{chen2013iboat, zhang2020continuous}. 
For example, the approach in~\cite{chen2013iboat} is to use some manually defined parameters to isolate an anomalous subtrajectory. It uses an adaptive window maintaining the latest incoming GPS points to compare against normal trajectories, where the normal trajectories are supported by the majority of the trajectories within an SD pair. As a new incoming point is added to window, it then checks the support of the subtrajectory in the window. If the support is larger than a pre-defined threshold, the point is labeled to be normal; otherwise, the point is labeled to be anomalous, and the adaptive window is reduced to contain only the latest point. The process continues until the trajectory is completed. The anomalous subtrajectory is recognized as those points labeled to be anomalous. However, the threshold is hard to set appropriately, and not general enough to cover all possible situations or adaptive to different road conditions.
{\Revision{Another approach in~\cite{zhang2020continuous} uses trajectory similarity (e.g., discrete Frechet) to detect anomalies. It computes the distance between a given normal trajectory and the current partial trajectory for each new incoming point, it reports an anomaly event if the distance exceeds a given threshold; otherwise, the detection continues.}}

}}

(3) \textbf{Requirement of sufficient supervision labels.}
While several machine learning models have been developed to identify the anomalous trajectories~\cite{song2018anomalous,li2007roam}, the effectiveness of those methods largely relies on sufficient labeled data under a supervised learning framework. Nevertheless, practical scenarios may involve a very limited amount of labeled anomalous trajectories compared with the entire trajectory dataset due to the requirement of heavy human efforts, \eg, the trajectory data used in~\cite{chen2013iboat} only contains 9 SD pairs. \\\vspace{-0.15in}

In this paper, we propose a novel reinforcement learning based solution \texttt{RL4OASD}, which avoids the aforementioned issues of existing approaches. First, \texttt{RL4OASD} is designed to detect anomalous \emph{subtrajectories} in an online fashion, and thus it avoids the first issue. It achieves this by predicting a normal/anomalous label for each road segment in a trajectory and detecting anomalous subtrajectories based on the labels of road segments. Second, \texttt{RL4OASD} is a data-driven approach relying on data \emph{without labels}, and thus it avoids the second and third issues. 
%
%
%
Specifically, \texttt{RL4OASD} consists of three components, namely data preprocessing, network RSRNet and network ASDNet (see Figure~\ref{fig:rl4oasd}). 
In data preprocessing, 
it conducts map-matching on raw trajectories, mapping them onto road networks and obtains the map-matched trajectories. 
Based on the map-matched trajectories, it computes some labels of the road segments involved in trajectories using some heuristics based on the historical transition data among road segments.
The labels, which are noisy, will be used to train
RSRNet in a weakly supervised manner. 
%
In RSRNet, 
it learns the representations of road segments based on both traffic context features and normal route features. These representations in the form of vectors will be fed into ASDNet to define the states of a Markov decision process (MDP) for labeling anomalous subtrajectories. 
%
In ASDNet, 
it models the task of labeling anomalous subtrajectories as an MDP and learns the policy via a policy gradient method~\cite{silver2014deterministic}. 
ASDNet outputs refined labels of road segments, which are further used to train RSRNet again, and then RSRNet provides better observations for ASDNet to train better policies. 
%
We emphasize that the noisy labels computed by the preprocessing component only provide some prior knowledge to address the cold-start problem of training RSRNet, but are not used to train the model as in a supervised paradigm.
%
Furthermore, the concept of anomalous trajectory might drift over time. For example, due to some varying traffic conditions (e.g., some accidents), the trajectory might gradually deviate from a normal route, and the concept of ``normal'' and ``anomalous'' is changed accordingly. \texttt{RL4OASD} can handle ``concept drift'' with an online learning strategy, i.e., it continues to be refined when new data comes in. 

Our contributions can be summarized as follows.
\begin{itemize}
    \item We propose the first deep reinforcement learning based solution to detect anomalous subtrajectories. The proposed model 1) can detect anomalous subtrajectories naturally, 2) is data-driven, and 3) does not require labeled data. 
    {\Revision{In addition, it can handle the concept drift of anomalous trajectories via online learning.}}
    
    \item We conduct extensive experiments on two real-world trajectory datasets, namely Chengdu and Xi'an. We compare our proposed solution with various baselines, and the results show that our solution is effective (e.g., it yields 20-30\% improvement compared to the best existing approach) and efficient (e.g., it takes less than 0.1ms to process each newly generated data point).
    
    \item {
     We manually label the anomalous subtrajectories for two real datasets, Chengdu and Xi'an, for testing. Each labeled dataset covers 200 SD pairs and 1,688 (resp. 1,057) map-matched trajectories with these SD pairs for the Chengdu dataset (resp. the Xi'an dataset). This labeled dataset is more than 50 times larger than the existing known dataset~\cite{chen2013iboat}. 
     \Revision{The manually labeled test data is publicly accessible via the link~\footnote{https://github.com/lizzyhku/OASD}.
    We believe this relatively large labeled dataset would help with comprehensive and reliable evaluations on approaches for anomalous subtrajectory detection.
    }}
\end{itemize}



\section{RELATED WORK}
\label{sec:related}


\subsection{Online Anomalous {\Revision{Trajectory/Subtrajectory}} Detection.}
Online anomalous trajectory detection aims to detect an ongoing trajectory in an online manner. Existing studies propose many methods for the problem and involve two categories: heuristic-based methods~\cite{chen2013iboat,zhang2020continuous} and learning-based methods~\cite{wu2017fast,liu2020online}.

For heuristic methods, Chen et al.~\cite{chen2013iboat} investigate the detection of anomalous trajectories in an online manner via the isolation-based method, which aims to check which parts of trajectories are isolated from the reference (i.e., normal) trajectories with the same source and destination routes. This method aims to detect the anomalous subtrajectories from ongoing trajectories, which is similar to our paper. However, this method models reference trajectories based on many manually-set parameters. Besides, the performance of this method is evaluated on a small manually labeled dataset with 9 SD pairs only, which fails to reflect the generalization of this method.
%
A recent work ~\cite{zhang2020continuous} proposes to calculate the trajectory similarity via discrete Frechet distance between a given reference route and the current partial route at each timestamp. If the deviation exceeds a given threshold at any timestamp, and the system alerts that an anomaly event of detouring is detected. There are many predefined parameters involved in this method. 
Our work differs from these heuristic-based studies in that it is based on a policy learned via reinforcement learning instead of the hand-crafted heuristic with many predefined parameters (e.g., the deviation threshold) for detecting anomalies. Besides, we evaluate the performance of our method on a large dataset.

For learning-based methods, a recent study~\cite{liu2020online} proposes to detect anomalous trajectories via a generation scheme, which utilizes the Gaussian mixture distribution to represent different kinds of normal routes and detects those anomalous trajectories that cannot be well-generated based on the given representations of normal routes. This method aims to detect whether the ongoing trajectory is anomalous or not. Another learning-based method~\cite{wu2017fast} proposes a probabilistic model to detect trajectory anomalies via modeling the distribution of driving behaviours from historical trajectories. The method involves many potential features that are associated with driving behaviour modeling, including road level and turning angle, etc. This method also only targets whether the online trajectory is anomalous or not. 

In our work, we target a finer-grained setting, i.e, detecting which part of an anomalous trajectory, namely subtrajectory, is responsible for its anomalousness in an online manner. Nevertheless, anomalous subtrajectory detection is not the focus in these studies.

\subsection{Offline Anomalous Trajectory Detection.}
Offline anomalous trajectory detection refers to detecting an anomalous trajectory or anomalous subtrajectories of a trajectory, where the trajectory inputted in an offline manner. Also, there are two types of studies. One type is heuristic-based methods~\cite{lee2008trajectory, zhang2011ibat,zhu2015time,lv2017outlier,banerjee2016mantra} and another type is learning-based methods~\cite{gray2018coupled,song2018anomalous}. {\Revision{For heuristic-based methods, some heuristic metrics involving distance or density metrics are usually utilized to detect anomalous trajectories. For example, an early study ~\cite{lee2008trajectory} proposes a partition and detect mechanism to conduct anomalous subtrajectory detection. The main idea is to partition the trajectories and detect the anomalous trajectory segments by computing the distance of each segment in a target trajectory to the segments in other trajectories. A recent study~\cite{lv2017outlier} utilizes edit distance metrics to detect anomalous trajectories based on mining the normal trajectory patterns. Another study~\cite{zhang2011ibat} proposes to cluster trajectories based on the same itinerary and further detect anomalous trajectories via checking whether the target trajectory is isolated from the cluster. In addition, ~\cite{ge2011taxi} proposes a system which is utilized to detect fraud taxi driving. The main idea of this method is to detect anomalous trajectories via combining distance and density metrics. 

Other studies are proposed to detect anomalous trajectories with learning-based methods. For example, Song et al~\cite{song2018anomalous} adopt recurrent neural network (RNN) to capture the sequential information of trajectories for detection. However, the proposed model needs to be trained in a supervised manner, and the labeled data is usually unavailable in real applications. Another learning-based work~\cite{gray2018coupled} aims at detecting anomalous trajectories by new moving subjects (i.e., new taxi driver detection) with an adversarial model (i.e., GAN) is adapted.}}
{\Comment{It is clear that these methods proposed in this line of research cannot be utilized for the online detection scenario. 
}}

\subsection{Other Types of Anomalous Detection Studies.}
Some studies~\cite{banerjee2016mantra,li2009temporal, ge2010top, yu2014detecting, zhu2018sub} focus on other types of anomalous trajectory detection, which are related to ours. We review them as follows. Banerjee et al.~\cite{banerjee2016mantra} study temporal anomaly detection, which employs travel time estimation and traverses some potential subtrajectories in a target trajectory, where the anomalies are identified if the travel time largely deviates from the expected time. Li et al.~\cite{li2009temporal} detect temporal anomalies and a method utilizing historical similarity trends is studied. In addition, Ge et al.~\cite{ge2010top} investigate the Top-K evolving anomalies.~\cite{yu2014detecting, zhu2018sub} detect anomalous trajectories in large-scale trajectory streams.

{\addComment{\subsection{Deep Reinforcement Learning.}
Deep reinforcement learning aims to guide an agent to make sequential decisions to maximize a cumulative reward, as the agent interacts with a specific environment, which is usually modeled as a Markov decision process (MDP)~\cite{puterman2014markov}. In recent years, reinforcement learning has attracted much research attention. For example, Oh et al.~\cite{oh2019sequential} explores inverse reinforcement learning for sequential anomaly detection, and Huang et al.~\cite{huang2018towards} designs a deep RL-based anomaly detector for time series detection. Wang et al.~\cite{wang2020efficient,wang2021similar} proposes to use RL-based algorithms to accelerate subtrajectory or sub-game similarity search. In addition, RL-based solutions are developed to simplify trajectories with different objectives~\cite{wang2021trajectory,wang2021error}.
In this paper, we propose a novel RL-based solution for online anomalous subtrajectory detection (called \texttt{RL4OASD}), and a policy gradient method~\cite{sutton2000policy} is adopted for solving the problem. To our best knowledge, this is the first deep reinforcement learning based solution for online anomalous subtrajectories detection.}}


\begin{figure*}
  \centering
  \includegraphics[width=0.98\linewidth]{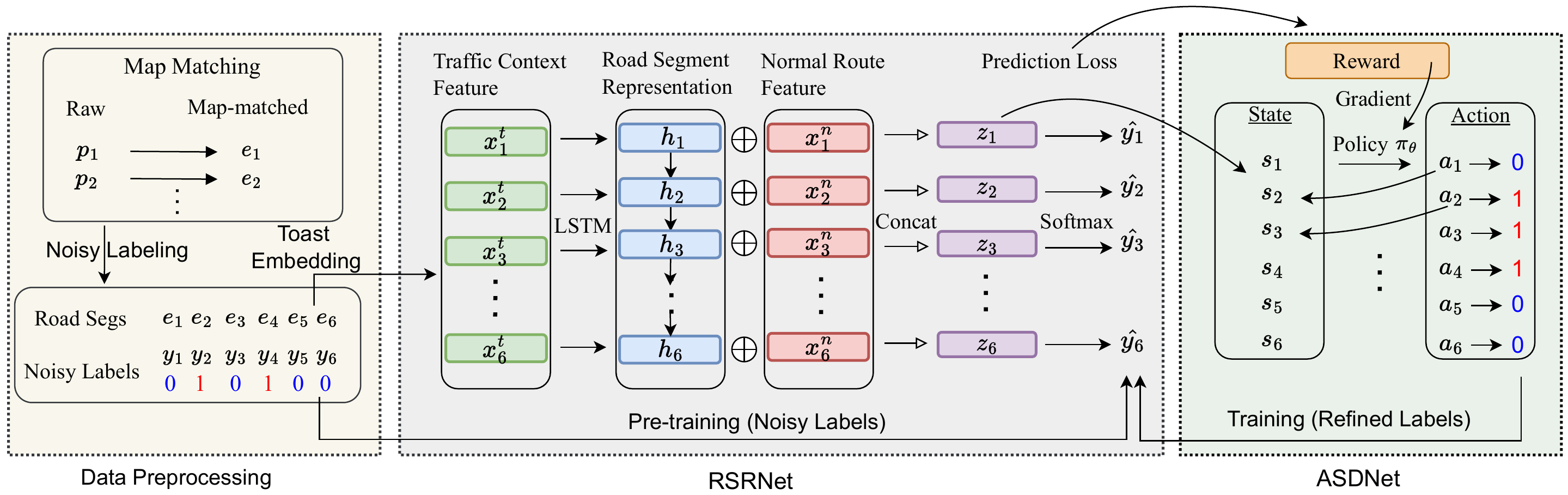}
  \vspace{-2mm}
  {\ICDERevision{\caption{Overview of \texttt{RL4OASD}, where $\bigoplus$ denotes the concatenation operation.}
  \label{fig:rl4oasd}
  }}
  \vspace{-4mm}
\end{figure*}

\section{PRELIMINARIES AND PROBLEM STATEMENT}
\label{sec:problem}

\subsection{Preliminaries}
\noindent \textbf{Raw Trajectory.} A raw trajectory $T$ consists of a sequence of GPS points, i.e., $T = <p_1, p_2,...,p_n>$, where a GPS point $p_i$ is in the form of a triplet, i.e., $p_i = (x_i, y_i, t_i)$, meaning that a moving object is located on $(x_i, y_i)$ at timestamp $t_i$, and $n$ represents the length of the trajectory $T$.

\smallskip
\noindent \textbf{Road Network.} A road network is represented as a directed graph $G(V,E)$, where $V$ represents a vertex set referring to crossroads or intersections, and $E$ represents an edge set referring to road segments, and for each edge $e$, it connects two vertexes $u$ and $v$ on the road network, denoted as $e=(u,v)$.

\smallskip
\noindent \textbf{Map-matched Trajectory.} A map-matched trajectory refers to a trajectory generated by a moving object on road networks, which corresponds to a sequence of road segments~\cite{yang2018fast}, i.e., $T = <e_1, e_2,...,e_n>$. 

For simplicity, we use $T$ to denote a map-matched trajectory, and we refer to map-matched trajectories as trajectories or routes interchangeably in the rest of the paper.

\smallskip
\noindent \textbf{Subtrajectory and Transition.} A subtrajectory $T[i,j]$ corresponds to a portion of the trajectory $T = <e_1, e_2,...,e_n>$ from $e_i$ to $e_j$, $1 \leq i \leq j \leq n$. A transition is defined as a special case of a subtrajectory, which only consists of two adjacent road segments,
i.e., $<e_{i-1}, e_{i}>$, where $1< i \leq n$.

\subsection{Problem Definition}
We study the problem of \emph{online anomalous subtrajectory detection}. Consider a source $S$ and destination $D$ pair (SD pair) and a set of trajectories $\mathcal{T}$ between them. Intuitively, a trajectory $T$ can be considered as \emph{normal} if it follows the route traveled by the majority of trajectories in $\mathcal{T}$. Based on this, we denote an \emph{anomalous} subtrajectory representing a part of a trajectory, which \emph{does not} follow the normal routes within the SD pair. We formulate the problem as follows.

\begin{problem}[OASD] Given an ongoing trajectory $T=<e_1,e_2,...e_n>$ that is generated from its source $S_T$ to destination $D_T$ in an online fashion, where the points $e_i$ are generated one by one, and the future points are not accessible in advance.
{\Revision{The OASD problem is to detect and update which parts of $T$ (i.e., subtrajectories) are anomalous, while T is sequentially generated.
}}
\label{prob:oasd}
\end{problem}

\section{METHODOLOGY}
\label{sec:method}

\subsection{Overview of \texttt{RL4OASD}}
\label{sec:overview_rl4oasd}

Determining whether a subtrajectory of an ongoing trajectory is anomalous or not is a decision-making process, which could be modeled as a Markov decision process (MDP)~\cite{puterman2014markov}. We propose a weakly supervised framework called \texttt{RL4OASD} (see Figure~\ref{fig:rl4oasd} for an overview).
The framework consists of three components, namely data preprocessing (Section~\ref{sec:weakly}), RSRNet (Section~\ref{sec:RSRNet}) and ASDNet (Section~\ref{sec:ASDNet}). 

In data preprocessing, 
we conduct map-matching on raw trajectories, mapping them onto road networks and obtain the map-matched trajectories. 
Based on the map-matched trajectories, we compute some labels of the road segments involved in trajectories using some heuristics based on the historical transition data among road segments.
The labels, which are noisy, will be used to train
RSRNet in a weakly supervised manner. 
%
In RSRNet, 
we learn the representations of road segments based on both traffic context features and normal route features. These representations in the form of vectors will be fed into ASDNet to define the states of the MDP for labeling anomalous subtrajectories. 
%
In ASDNet, 
we model the task of labeling anomalous subtrajectories as a MDP and learn the policy via a policy gradient method~\cite{silver2014deterministic}. 
ASDNet outputs refined labels of road segments, which are further used to train RSRNet again, and then RSRNet provides better observations for ASDNet to train better policies. The process iterates and we call the resulting algorithm combining RSRNet and ASDNet as \texttt{RL4OASD} (Section~\ref{sec:RL4OASD}).

{\ICDERevision{We explain some insights behind the effectiveness of \texttt{RL4OASD} as follows. First, in RSRNet, 
both traffic context features (e.g., driving speed, trip duration) and normal route features that are associated with the anomalies are well captured into the model. Second, in ASDNet, the task of labeling anomalous subtrajectories is formulated as a MDP, whose policy is learned in a data-driven manner, instead of using heuristics (e.g., some pre-defined parameters) as the existing studies do.
%
Third, 
{\CHENG RSRNet is first trained with noisy labels for dealing with the cold-start problem.
Then, RSRNet and ASDNet are trained iteratively and collaboratively, where RSRNet uses ASDNet's outputs as labels and ASDNet uses the outputs of RSRNet as features.}
}}

\subsection{Data Preprocessing}
\label{sec:weakly}
The data processing component involves a map matching process~\cite{yang2018fast} and a process of obtaining noisy labels.
The latter produces some noisy labels, which will be utilized to pre-train the representations in RSRNet and also to provide a warm-start for the policy learning in ASDNet.
Recall that we do not assume the availability of real labels for training since manually labeling the data is time-consuming.
Specifically, the process of obtaining noisy labels involves four steps. 

\underline{Step-1}: 
We group historical trajectories in a dataset with respect to different SD pairs and time slots. Here, we have 24 time slots if we partition one day with one hour granularity, and we say a trajectory falls into a time slot if its starting travel time is within the slot. For example, in Figure~\ref{fig:example}, we have three map-matched trajectories $T_1$, $T_2$ and $T_3$ with the source $e_1$ and destination $e_{10}$. Assuming the starting travel times of $T_1$, $T_2$ and $T_3$ are 9:00, 9:10, and 9:30, respectively. Then, all three trajectories are in the same group because they are within the same time slot with one hour granularity.

\underline{Step-2}: We then compute the fraction of \emph{transitions} each from a road segment to another with respect to all trajectories in each group. Suppose that there are 5 trajectories traveling along $T_1$, 4 along $T_2$, and only 1 along $T_3$, the fraction of \emph{transition} $<e_1,e_2>$ is calculated as $5/10=0.5$ since it appears 5 times (along $T_2$ and $T_3$) in all 10 trajectories.

\underline{Step-3}: For each trajectory, we refer to the group it belongs to and map it to a sequence of \emph{transition} fractions with respect to each road segment. For example, for a trajectory traveling the route as $T_3$, its mapped \emph{transition} sequence is $<<*,e_1>, <e_1,e_2>, <e_2,e_4>, <e_4, e_{11}>, <e_{11}, e_{12}>, <e_{12}, e_{13}>, <e_{13}, e_{14}>, <e_{14}, e_{15}>, <e_{15}, e_{10}>>$, where we pad the initial transition as $<*,e_1>$, and the corresponding fraction sequence is $<1.0,0.5,0.5,0.1,0.1,0.1,0.1,0.1,1.0>$. Note that the fractions on the source $e_1$ (corresponding to $<*,e_1>$) and the destination $e_{10}$ (corresponding to $<e_{15}, e_{10}>$) are always set to 1.0, since the source and destination road segments are definitely travelled within its group.

\underline{Step-4}: We obtain the noisy labels by using a threshold parameter $\alpha$, where 0 denotes a normal road segment, whose fraction is larger than $\alpha$ meaning that the road segment is frequently traveled, and 1 otherwise. For example, by using the threshold $\alpha=0.5$, we obtain the noisy labels of $T_3$ as $<0,1,1,1,1,1,1,1,0>$.


\subsection{Road Segment Representation Network (RSRNet)}
\label{sec:RSRNet}
In RSRNet, we adopt the LSTM~\cite{hochreiter1997long} structure, which accepts trajectories with different lengths and captures the sequential information behind trajectories. We embed two types of features into representations, namely the traffic context features on road networks and the normal route features for a given SD pair.  

\smallskip
\noindent \textbf{Traffic Context Feature (TCF).} A map-matched trajectory corresponds to a sequence of road segments and each road segment is naturally in the form of a token (i.e., the road segment id). We pre-train each road segment in the embedding layer of RSRNet as a vector, which captures traffic context features (e.g., driving speed, trip duration, road type). To do this, we employ Toast~\cite{chen2021robust}, which is a recent road network representation learning model to support road segment based applications. The learned road segment representations will be used to initialize the embedding layer in RSRNet, and those embeddings can be further optimized with the model training. Other road network representation models~\cite{jepsen2019graph,wang2019learning} are also applicable for the task.

\smallskip
\noindent \textbf{Normal Route Feature (NRF).} Given an SD pair in a time slot, we first infer the normal routes within it. Intuitively, the normal trajectories often follow the same route. If a trajectory contains some road segments that are rarely traveled by others, it probably contains anomalies. Therefore, we infer a route as the normal route by calculating the fraction of the trajectories passing through the route with respect to all trajectories within its SD pair. The inferred results are obtained via comparing a threshold (denoted by $\delta$) with the fraction of each road segment, i.e., a route is inferred as the normal if the fraction is larger than the threshold, and vice versa.
{\ICDERevision{For example, in Figure~\ref{fig:example}, recall that there are 5 trajectories traveling along $T_1$, 4 along $T_2$, and only 1 along $T_3$.
Given $\delta=0.3$, we infer $T_1$ (and resp. $T_2$) as the normal route, because its fraction $5/10=0.5$ (and resp. $4/10=0.4$) is larger than the threshold $\delta$ in all 10 trajectories.}}
%
%
Based on the inferred normal routes, we then extract the features of a trajectory. For example, given a trajectory following $T_3$, 
we extract the features as follows: a road segment in a target trajectory is normal (i.e., 0) if the transition on that road segment occurs in the inferred normal routes; and 1 otherwise. For example, the extracted normal features of $T_3$ are $<0,0,0,1,1,1,1,1,0>$, where the feature on the road segment $e_2$ is 0 since the transition $<e_1,e_2>$ occurs in the normal route $T_2$. Note that the source and destination road segments always have the feature 0 (i.e., normal).
{\Revision{We notice normal route features and noisy labels are both in the form of 0-1. The difference between them is that the former is to capture the information of normal routes, and correspondingly it is obtained by normal routes at a route-level. In contrast, the latter is used as the labels for training RSRNet, which is obtained by computing transition frequencies at an edge-level. {\ICDERevision{In addition, the former is utilized during the whole training process, while the latter is utilized to pre-train the representations in RSRNet only (which would provide a warm-start for ASDNet).}}
}}
After obtaining the normal route features that are in the form of tokens, we then obtain a vector for the feature by embedding the tokens as one-hot vectors. 
We call the obtained vectors embedded normal route features.

\smallskip
\noindent \textbf{Training RSRNet.} Figure~\ref{fig:rl4oasd} illustrates the architecture of RSRNet. In particular, given a sequence of embedded traffic context features $\mathbf{x_i}^{t}$ ($1 \leq i \leq n$), where $n$ denotes the trajectory length, the LSTM obtains the hidden state $\mathbf{h_i}$ at each road segment. 
We then concatenate $\mathbf{h_i}$ with the embedded normal route feature $\mathbf{x_i}^{n}$, denoted by $\mathbf{z}_i=[\mathbf{h}_i;\mathbf{x}_i^{n}]$. 
Note that the two parts $\mathbf{h}$ and $\mathbf{x}^{n}$ capture the sequential trajectory information and normal route information, respectively. 
Note that we do not let $\mathbf{x}^{n}$ go through the LSTM since it preserves the normal route feature at each road segment.
We adopt cross-entropy loss to train RSRNet between the predicted label $\hat{y}_i$ based on the $\mathbf{z}_i$ and the {\Revision{noisy/refined label}} $y_i$, i.e.,
\begin{equation}
\label{eq:entropy}
\mathcal{L}=\frac{1}{n}\sum_{i=1}^{n}\mathcal{H}(y_i,\hat{y}_i),
\end{equation}
where $\mathcal{H}$ denotes the cross-entropy operator. 

\subsection{Anomalous Subtrajectory Detection Network (ASDNet)}
\label{sec:ASDNet}
Consider the task of online anomalous subtrajectory detection is to sequentially scan an ongoing trajectory, and for each road segment, decide whether an anomaly happened at that position. This motivates us to model the process as a Markov decision process (MDP), involving states, actions, 
and rewards. 

\smallskip
\noindent \textbf{States.} We denote the state when scanning the road segment $e_i$ as $\mathbf{s}_i$. The state $\mathbf{s}_i$ ($1 < i \leq n$) is represented as the concatenation of $\mathbf{z}_i$ and $\mathbf{v}(e_{i-1}.l)$, i.e., $\mathbf{s}_i=[\mathbf{z}_i;\mathbf{v}(e_{i-1}.l)]$, where $\mathbf{z}_i$ is obtained from RSRNet and $\mathbf{v}(e_{i-1}.l)$ denotes a vector of the label on the previous road segment $e_{i-1}$ by embedding its token (i.e., 0 or 1). The rationale for the state design is to capture the features from three aspects, i.e., traffic context, normal routes and the previous label.

\smallskip
\noindent \textbf{Actions.} We denote an action of the MDP by $a$, which is to label each road segment as normal or not. Note that an anomalous subtrajectory boundary can be identified when the labels of two adjacent road segments are different.

\smallskip
\noindent \textbf{Rewards.} The reward involves two parts. One is called \emph{local reward}, which aims to capture the local continuity of labels of road segments. The rationale is that the labels of normal road segments or anomalous road segments would not change frequently. The second one is called \emph{global reward}, which aims to indicate the quality of the refined labels
(indicated by the cross-entropy loss of RSRNet). We take the loss as some feedback to guide the training of ASDNet.


The local reward is an intermediate reward, which encourages the continuity of the labels on road segments. Specifically, it is defined as 
\begin{equation}
r_i^{local}=\text{sign}(e_{i-1}.l=e_{i}.l)\cdot \text{cos}(\mathbf{z}_{i-1},\mathbf{z}_{i}),
\end{equation}
where the $\text{sign}(e_{i-1}.l=e_{i}.l)$ returns 1 if the condition $e_{i-1}.l=e_{i}.l$ is true {\Revision{(i.e., the labels are continued)}} and -1 otherwise; $\text{cos}(\mathbf{z}_{i-1},\mathbf{z}_{i})$ denotes the consine similarity between $\mathbf{z}_{i-1}$ and $\mathbf{z}_{i}$, which are obtained from RSRNet. {\Comment{We choose the consine similarity because it outputs a normalized value between 0 and 1, which corresponds to the same output range (i.e., between 0 and 1) of the global reward.
}}

The global reward is designed to measure the quality of refined labels by ASDNet. We feed the refined labels into RSRNet, and compute the global reward as 
\begin{equation}
    r^{global}= \frac{1}{1+\mathcal{L}},
\end{equation}
where $\mathcal{L}$ denotes the cross-entropy loss in Equation~\ref{eq:entropy}. We notice this reward has the range between 0 and 1, and makes the objective of RSRNet well aligned with ASDNet (a smaller loss $\mathcal{L}$ means a larger global reward).

\smallskip
\noindent \textbf{Policy Learning on the MDP.} 
The core problem of a MDP is to learn a policy, which guides an agent to choose actions based on the constructed states such that the cumulative reward, denoted by $R_n$, is maximized. We learn the policy via a policy gradient method~\cite{silver2014deterministic}, called the REINFORCE algorithm~\cite{williams1992simple}. To be specific, let $\pi_{\theta}(a|\mathbf{s})$ denote a stochastic policy, which is used to sample an action $a$ for a given state $\mathbf{s}$ via a neural network, whose parameters are denoted by $\theta$. Then, the gradients of some performance measure $J(\theta)$ wrt the network parameters $\theta$ are estimated as
\begin{equation}
\label{eq:policy}
\nabla_{\theta}J(\theta) = \sum_{i=2}^{n}R_n\nabla_{\theta}\ln\pi_{\theta}(a_i|\mathbf{s}_i),
\end{equation}
which can be optimized using an optimizer (e.g., Adam stochastic gradient ascent). The expected cumulative reward $R_n$ for a trajectory is defined as
\begin{equation}
R_n=\frac{1}{n-1}\sum_{i=2}^{n}r_i^{local}+r^{global}.
\end{equation}

\smallskip
\noindent \textbf{Joint Training of RSRNet and ASDNet.} We jointly train the two networks (e.g., RSRNet and ASDNet). First, we map raw trajectories on road networks and generate the noisy labels as explained in Section~\ref{sec:weakly}. Then, we randomly sample 200 trajectories to pre-train the RSRNet and ASDNet, separately. In particular, for the RSRNet, we train the network in a supervised manner with the noisy labels. For the ASDNet, we specify its actions as the noisy labels, and train the policy network via a gradient ascent step as computed by Equation~\ref{eq:policy}. The pre-training provides a warm-start for the two networks, where the parameters of the two networks have incorporated some information captured from the normal routes before joint training.

During the joint training, we randomly sample 10,000 trajectories, and for each trajectory, we generate 5 epochs to iteratively train the RSRNet and ASDNet. In particular, we apply the learned policy from ASDNet to 
obtain refined labels, and the refined labels are used to train the RSRNet to obtain the better representation $\mathbf{z}_i$ on each road segment. Then, $\mathbf{z}_i$ is used to construct the state to learn a better policy in ASDNet. Since the learned policy can further refine the labels, and the two networks are jointly trained with the best model is chosen during the process.

\subsection{The \texttt{RL4OASD} Algorithm}
\label{sec:RL4OASD}
\begin{algorithm}[t!]
	\caption{The \texttt{RL4OASD} algorithm}
	\label{alg:rl4oasd}
		\KwIn{
		A map-matched trajectory $T = <e_1, e_2, ..., e_n>$ which is inputted in an online manner}
        \For{$i=1,2,...,n$}{
			\uIf{$i=1$ \textbf{or} $i=n$}{
			$e_{i}.l \leftarrow 0$;
			}
           \uElse{
           Call RSRNet for obtaining a representation $\mathbf{z}_{i}$;\\
		   Construct a state $\mathbf{s}_i=[\mathbf{z}_i;\mathbf{v}(e_{i-1}.l)]$;\\
		   Sample an action (i.e., 0 or 1), $a_i \sim \pi_{\theta}(a|\mathbf{s})$;\\
		   $e_{i}.l \leftarrow a_i$;\\
		   {\Revision{Monitor the anomalous subtrajectory consisting of the road segments with the label 1 and return the subtrajectory when it is formed;}}
        }
}
{\Revision{\textbf{Return} a NORMAL trajectory signal;}}
\end{algorithm}

Our \texttt{RL4OASD} algorithm is based on the learned policy for detecting anomalous subtrajectories. The process is presented in Algorithm~\ref{alg:rl4oasd}. Specifically, \texttt{RL4OASD} accepts an ongoing trajectory in an online fashion for labeling each road segment as the normal (i.e., 0) or not (i.e., 1) (lines 1-9). It first labels the source or destination road segment as normal by definitions (lines 2-3). It then 
constructs a state via calling RSRNet (lines 5-6), and samples an action to label the road segment based on the learned policy in ASDNet (lines 7-8). {\Revision{It monitors the anomalous subtrajectory that involves the abnormal labels (i.e., 1) on the road segments 
and returns it when it is formed (line 9). Finally, if no anomaly is detected, the algorithm returns a NORMAL trajectory signal (line 11).}} We further develop two enhancements for boosting the effectiveness and efficiency of \texttt{RL4OASD}. One is called Road Network Enhanced Labeling (RNEL), and the other is called Delayed Labeling (DL).

\smallskip
\noindent\textbf{Road Network Enhanced Labeling.}
{\Comment{For the RNEL, we utilize the graph structure of a road network to help labeling a road segment, where a label on a road segment is deterministic in one of three cases: (1) If $e_{i-1}.out=1$, $e_{i}.in=1$, then $e_{i}.l=e_{i-1}.l$. (2) If $e_{i-1}.out=1$, $e_{i}.in>1$ and $e_{i-1}.l=0$, then $e_{i}.l=0$. (3) If $e_{i-1}.out>1$, $e_{i}.in=1$ and $e_{i-1}.l=1$, then $e_{i}.l=1$. Here, $e_{i}.out$ and $e_{i}.in$ denote the out degree and in degree of a road segment $e_{i}$, respectively, and $e_{i}.l$ denotes the label of the road segment $e_{i}$. 
The common intuition behind them is that: (a) any change of the anomalousness status from normal (0) at $e_{i-1}$ to abnormal (1) at $e_i$ means that there exist alternative transitions from $e_{i-1}$ to other road segments (i.e., $e_{i-1}.out > 1$); and (b) any change of the anomalousness status from abnormal (1) at $e_{i-1}$ to normal (0) at $e_i$ means that there exist alternative transitions from other road segments to $e_{i}$ (i.e., $e_{i}.in > 1$).
Based on the rules, we only perform actions in the otherwise cases via the RL model, and some potential wrong decisions can therefore be avoided. In addition, the efficiency can be improved since the time of taking actions in some cases is saved via checking the rules instead of calling the RL model.
}}

\smallskip
\noindent\textbf{Delayed Labeling.}
For the DL, \texttt{RL4OASD} forms an anomalous subtrajectory whenever its boundary is identified, i.e., the boundary is identified at $e_{i-1}$ if $e_{i-1}.l = 1$ and $e_{i}.l=0$. Intuitively, it looks a bit rush to form an anomalous subtrajectory and may produce many short fragments from a target trajectory. 
Therefore, we consider a delay technique as a post-processing step. Specifically, it scans $D$ more road segments that follow $e_{i-1}$ when forming an anomalous subtrajectory. Among the $D$ road segments, we choose the final position $j$ ($i-1<j \leq i-1+D$) where the road segment is with label 1, and then convert some 0's to 1's between the position $i-1$ and $j$.
It could be verified the labeling with delay does not incur much time cost, and offers a better continuity to avoid forming too many fragments.

\smallskip
\noindent \textbf{Time complexity.} The time complexity of the \texttt{RL4OASD} algorithm is $O(n)$, where $n$ denotes the length of a target trajectory. The time is dominated by two networks, i.e., RSRNet and ASDNet. We analyze them as follows. In RSRNet, the time cost for one road segment consists of (1) that of obtaining embeddings of TCF and NRF, which are both $O(1)$ and (2) that of obtaining the $\mathbf{z}$ via a classic LSTM cell, which is $O(1)$. In ASDNet, the time cost for one road segment consists of (1) that of constructing a state, where the part $\mathbf{z}$ has been computed in RSRNet and the part $\mathbf{v}(e.l)$ is obtained via an embedding layer, which is $O(1)$ and (2) that of sampling an action via the learned policy, which is $O(1)$. As we can see in Algorithm~\ref{alg:rl4oasd}, the two networks are called at most $n$ times. Therefore, the time complexity of the \texttt{RL4OASD} is $O(n)\times O(1)=O(n)$. We note that the $O(n)$ time complexity maintains the current best time complexity as those of existing algorithms~\cite{chen2013iboat,wu2017fast,liu2020online} for the trajectory or subtrajectory anomaly detection task, and can largely meet practical needs for online scenarios as shown in our experiments.

\smallskip
{\Revision{\noindent \textbf{Handling Concept Drift of Anomalous Trajectories.} As discussed in Section~\ref{sec:problem}, we detect anomalous subtrajectories that do not follow the normal routes. However, the concept of ``normal'' and ``anomalous'' may change over time with varying traffic conditions. For example, when some popular route is congested, then drivers may gradually prefer to travel another unpopular route (i.e., to avoid traffic jams). In this case, the unpopular route should be considered as a normal route, while the trajectories traveling the previous route may become anomalous. To handle the issue caused by the concept drift of the normal and anomalous, we adopt an online learning strategy~\cite{liu2020online}, where the model continues to train with newly recorded trajectory data, and keeps its policy updated for the current traffic condition (we have validated the effectiveness of the strategy in Section~\ref{sec:concept_exp}).
}}

\smallskip
{\ICDERevision{\noindent \textbf{Discussion on the cold-start problem.} As the anomalous subtrajectories are defined as unpopular parts, and thus there may exist a cold-start problem for the detection, where the historical trajectories are not sufficient for some SD pairs. 
Our method relies on the historical trajectories for defining the normal route feature. The feature is calculated as a relative fraction between 0 and 1. Specifically, the feature of a route is defined to be the number of trajectories along the route over the total number of trajectories within the SD pair.
{\CHENG We have conducted experiments} by varying the number of historical trajectories within the SD pairs in Table~\ref{tab:droprate}. {\CHENG The results show that our model is robust against sparse data, e.g., its effectiveness} only degrades by 6\% even if 80\% of historical trajectories are dropped. The cold-start problem in anomalous trajectory/subtrajectory detection looks very interesting. We believe that some generative methods, e.g., to generate some routes within the sparse SD pairs, can possibly be leveraged to overcome the issue, which we plan to explore as future work.
}}

\section{EXPERIMENTS}
\label{sec:experiment}
\subsection{Experimental Setup}
\label{sec:setup}

\noindent \textbf{Dataset.}
The experiments are conducted on two real-world taxi trajectory datasets from DiDi Chuxing~\footnote{https://outreach.didichuxing.com/research/opendata/en/}, namely \emph{Chengdu} and \emph{Xi'an}. All raw trajectories are preprocessed to map-matched trajectories via a popular map-matching algorithm~\cite{yang2018fast}, and the road networks of the two cities are obtained from OpenStreetMap~\footnote{https://www.openstreetmap.org/}. {\Revision{Following previous studies~\cite{liu2020online,chen2013iboat}, we preprocess the datasets and filter those SD-pairs that contain less than 25 trajectories to have sufficient trajectories to indicate the normal routes. We randomly sample 10,000 trajectories from the datasets for training, and the remaining for testing.
}}

\smallskip
\noindent \textbf{Ground Truth.} 
By following the previous works~\cite{zhang2011ibat,chen2013iboat, lv2017outlier}, we manually label the anomalous subtrajectories, and take the labeled subtrajectories as the ground truth for the evaluation.
{\Revision{Specifically, we sample 200 SD pairs with sufficient trajectories between them (e.g., at least 30 trajectories for each pair, and over 900 trajectories on average).}}
{\ICDERevision{For labeling subtrajectories, we illustrate all routes (i.e., map-matched trajectories) within each SD pair and highlight the road segments that are traveled by the majority of trajectories as shown in Figure~\ref{fig:casestudy}. We invite 5 participants, and first spend 10 minutes to get everyone to understand the visualization, then let them identify whether the routes are anomalous or not based on visual inspection. If a participant thinks the route is anomalous, we then ask the participant to label which parts are responsible for its anomalousness.
}} 
For quality control, we randomly pick 10\% trajectories, ask 5 other checkers to label these trajectories independently, adopt the majority voting to aggregate the labels, and compute the accuracy of the labels by the labelers against the aggregated ones by the checkers. The accuracy is 98.7\% for Chengdu and 94.3\% for Xi'an, which shows that our labeled datasets are with high accuracy.

Multiple raw trajectories may correspond to the same route and thus we have fewer routes than raw trajectories. We label 1,688 (resp. 1,057) routes, which correspond to 558,098 (resp. 163,027) raw trajectories before map-matching for Chengdu (resp. Xi'an). Among them, 1,436 (resp. 813) routes that correspond to 3,930 (resp. 2,368) raw trajectories are identified as the anomalous for Chengdu (resp. Xi'an), where the anomalous ratios for Chengdu and Xi'an are estimated as $3,930/558,098=0.7\%$ and $2,368/163,027=1.5\%$, respectively. The statistics of the datasets are reported in Table~\ref{tab:dataset}.

\begin{table}[]
\setlength{\tabcolsep}{8pt}
\centering
\vspace{-4mm}
\caption{Dataset statistics.}
\vspace{-3mm}
\begin{tabular}{lll}
\hline
Dataset             & Chengdu       & Xi'an         \\ \hline
\# of trajectories      & 677,492       & 373,054       \\
\# of segments          & 4,885         & 5,052         \\
\# of intersections     & 12,446        & 13,660        \\
\# of labeled routes (trajs)  & 1,688 (558,098) & 1,057 (163,027) \\
\# of anomalous routes (trajs) & 1,436 (3,930)   & 813 (2,368)     \\
Anomalous ratio     & 0.7\%         & 1.5\%         \\ 
Sampling rate & 2s $\sim$ 4s  &  2s $\sim$ 4s\\
\hline
\end{tabular}
\label{tab:dataset}
\vspace{-4mm}
\end{table}

\smallskip
\noindent \textbf{Baseline.}
We review the literature thoroughly, and identify the following baselines for the online detection problem, including IBOAT~\cite{chen2013iboat}, DBTOD~\cite{wu2017fast}, GM-VSAE~\cite{liu2020online}, SD-VSAE~\cite{liu2020online}, SAE~\cite{liu2020online}, VSAE~\cite{liu2020online} and CTSS~\cite{zhang2020continuous}. The detailed description of these algorithms are represented as follows:
{\addComment{
\begin{itemize}[leftmargin=*]
\item \textbf{IBOAT}~\cite{chen2013iboat}: it is an online method to detect anomalous trajectories via checking which parts of trajectories isolate from reference (i.e., normal) trajectories for the same source and destination routes.   
\item \textbf{DBTOD}~\cite{wu2017fast}: it utilizes a probabilistic model to perform anomalous trajectory detection by modeling human driving behaviors from historical trajectories. 
\item \textbf{GM-VSAE}~\cite{liu2020online}: it is a method aiming to detect anomalous trajectories via a generation scheme. This method utilizes the Gaussian mixture distribution to represent categories of different normal routes. Then based on these representations, the model detects anomalous trajectories which are not well-generated. 
\item \textbf{SD-VSAE}~\cite{liu2020online}: it is a fast version of GM-VSAE, which outputs one representation of the normal route with the maximized probability. And based on this normal route representation, the model detects anomalous trajectories that cannot be generated well followed by GM-VSAE.
\item \textbf{SAE}~\cite{liu2020online}: we adapt GM-VSAE, where SAE replaces the encoder and decoder structure in GM-VSAE with a traditional Seq2Seq model, which aims to minimize a reconstruction error, and the reconstruction error is further used to define the anomaly score.
\item \textbf{VSAE}~\cite{liu2020online}: we also adapt GM-VSAE by using VSAE to replace the Gaussian mixture distribution of the latent route representations in GM-VSAE with a Gaussian distribution.
\item \textbf{CTSS}~\cite{zhang2020continuous}: it is a method to detect anomalous trajectories via calculating the trajectory similarity with discrete Frechet distance between a given reference route and the current partial route at each timestamp. 
\end{itemize}
}}
{\Revision{We notice the baselines~\cite{wu2017fast,liu2020online,zhang2020continuous} are proposed for anomalous trajectory detection, and output an anomaly score on each point in a trajectory. We note that those scores are computed from the beginning, i.e., they only consider the subtrajectories starting from the source, which causes them difficult to be adapted for the subtrajectory detection task, where the detected subtrajectories can be started at any position of the trajectory. Thus, we adapt them in this way. 
We tune their thresholds of the anomaly scores in a development set (i.e, a set of 100 trajectories with manual labels), and for each tuned threshold, the anomalous subtrajectories are identified as those road segments, whose anomaly scores are larger than the threshold. The threshold that is associated with the best performance (evaluated by $F_1$-score and details will be presented later) is selected for experiments. In addition, we also tune the parameters of baselines to the best based on the development set.
For \texttt{RL4OASD}, it naturally outputs the anomalous subtrajectories for experiments. 
}}

\begin{table*}[]
\vspace*{-4mm}
{\ICDERevision{\caption{Effectiveness comparison with existing baselines (Left: $F_1$-score, right: $TF_1$-score).}
\label{tab:baselines}}}
\setlength{\tabcolsep}{2.4pt}
\vspace*{-3mm}
\centering
\begin{tabular}{|c|c|c|c|c|c|c|c|c|c|c|c|c|c|c|c|c|c|c|c|c|}
\hline
Methods           & \multicolumn{10}{c|}{Chengdu} & \multicolumn{10}{c|}{Xi'an} \\ \hline
\begin{tabular}[c]{@{}c@{}}Trajectory \\ Length\end{tabular}   & \multicolumn{2}{c|}{G1}    & \multicolumn{2}{c|}{G2}    & \multicolumn{2}{c|}{G3}    & \multicolumn{2}{c|}{G4}  &\multicolumn{2}{c|}{Overall} & \multicolumn{2}{c|}{G1}    & \multicolumn{2}{c|}{G2}   & \multicolumn{2}{c|}{G3}   & \multicolumn{2}{c|}{G4}  &\multicolumn{2}{c|}{Overall} \\ \hline
IBOAT~\cite{chen2013iboat}             &0.534 &{\ICDERevision{0.541}}       &0.538 &{\ICDERevision{0.544}}       &0.519 &{\ICDERevision{0.525}}      &0.674 &{\ICDERevision{0.781}}  &0.539 &{\ICDERevision{0.550}}  &0.556 &{\ICDERevision{0.615}}       &0.493 &{\ICDERevision{0.495}}      &0.497 &{\ICDERevision{0.548}}     &0.493 &{\ICDERevision{0.463}} &0.506 &{\ICDERevision{0.519}}   \\ \hline
DBTOD~\cite{wu2017fast}             &0.523 &{\ICDERevision{0.533}}
&0.531 &{\ICDERevision{0.537}}       &0.515 &{\ICDERevision{0.522}}      &0.669 &{\ICDERevision{0.750}} &0.530 &{\ICDERevision{0.542}}   &0.519 &{\ICDERevision{0.523}}      &0.448 &{\ICDERevision{0.381}}     &0.441 &{\ICDERevision{0.380}}     &0.483 &{\ICDERevision{0.452}} &0.433  &{\ICDERevision{0.424}}   \\ \hline
GM-VSAE~\cite{liu2020online}           &0.375 &{\ICDERevision{0.383}}       &0.493 &{\ICDERevision{0.498}}      &0.507 &{\ICDERevision{0.513}}     &0.669 &{\ICDERevision{0.750}} &0.452&{\ICDERevision{0.461}}  &0.270&{\ICDERevision{0.272}}     &0.361  &{\ICDERevision{0.308}}    &0.389  &{\ICDERevision{0.332}}    &0.421 &{\ICDERevision{0.388}} &0.363  &{\ICDERevision{0.328}}     \\ \hline
SD-VSAE~\cite{liu2020online}           &0.375 &{\ICDERevision{0.383}}     &0.463 &{\ICDERevision{0.466}}       &0.416 &{\ICDERevision{0.406}}      &0.555 &{\ICDERevision{0.612}} &0.452 &{\ICDERevision{0.461}}   &0.270 &{\ICDERevision{0.272}}      &0.353 &{\ICDERevision{0.300}}     &0.385 &{\ICDERevision{0.329}}     &0.386 &{\ICDERevision{0.360}}  &0.350  &{\ICDERevision{0.319}}  \\ \hline
SAE~\cite{liu2020online}               &0.375 &{\ICDERevision{0.383}}       &0.461 &{\ICDERevision{0.463}}      &0.413 &{\ICDERevision{0.406}}      &0.536 &{\ICDERevision{0.603}}  &0.451  &{\ICDERevision{0.461}}  &0.270 &{\ICDERevision{0.272}}   &0.359 &{\ICDERevision{0.300}}     & 0.386 &{\ICDERevision{0.329}}     &0.410 &{\ICDERevision{0.379}} &0.363 &{\ICDERevision{0.328}}     \\ \hline
VSAE~\cite{liu2020online}              &0.375 &{\ICDERevision{0.383}}      &0.491 &{\ICDERevision{0.496}}      &0.493 &{\ICDERevision{0.497}}      &0.655 &{\ICDERevision{0.734}} &0.448 &{\ICDERevision{0.457}}  &0.262  &{\ICDERevision{0.265}}      &0.340  &{\ICDERevision{0.296}}     &0.371  &{\ICDERevision{0.316}}     &0.369 &{\ICDERevision{0.345}} &0.339 &{\ICDERevision{0.309}}   \\ \hline
CTSS~\cite{zhang2020continuous}              &0.730 &{\ICDERevision{0.786}}      &0.708 &{\ICDERevision{0.764}}      &0.625 &{\ICDERevision{0.657}}      &0.741 &{\ICDERevision{0.845}} &0.706 &{\ICDERevision{0.758}}  &0.657 &{\ICDERevision{0.669}}      &0.637 &{\ICDERevision{0.688}}     &0.636 &{\ICDERevision{0.689}}     &0.672 &{\ICDERevision{0.700}} &0.658 &{\ICDERevision{0.689}}   \\ \hline
\texttt{RL4OASD}          &\textbf{0.888} &{\ICDERevision{\textbf{0.905}}}      &\textbf{0.892} &{\ICDERevision{\textbf{0.910}}}      &\textbf{0.725} &{\ICDERevision{\textbf{0.720}}} &\textbf{0.774} &{\ICDERevision{\textbf{0.853}}} &\textbf{0.854} &{\ICDERevision{\textbf{0.870}}}    &\textbf{0.964}  &{\ICDERevision{\textbf{0.973}}}     &\textbf{0.864} &{\ICDERevision{\textbf{0.888}}}     &\textbf{0.809} &{\ICDERevision{\textbf{0.843}}}    &\textbf{0.844} &{\ICDERevision{\textbf{0.870}}} &\textbf{0.857} &{\ICDERevision{\textbf{0.883}}}   \\ \hline
\end{tabular}
\vspace*{-4mm}
\end{table*}

\smallskip
\noindent \textbf{Parameter Setting.} In RSRNet, we embed the TCF and NRF features into the 128-dimensional vectors, and use the LSTM with 128 hidden units to implement the RSRNet. {\Revision{The parameter $\alpha$, $\delta$ and $D$ are set to 0.5, 0.4 and 8, respectively.}}
%
To train RSRNet, we compute noisy labels in 24 time slots with one hour granularity via empirical studies. In ASDNet, the dimension of label vectors $\mathbf{v}(e_{i}.l)$ is 128. The policy network is implemented with a single-layer feedforward neural network, and the softmax function is adopted as the activation function. By empirical findings, the learning rates for RSRNet and ASDNet are set to 0.01 and 0.001, respectively.

\begin{table}[t!]
\setlength{\tabcolsep}{10pt}
\centering
\caption{Ablation study for \texttt{RL4OASD}.}
\vspace{-3mm}
\begin{tabular}{|c|c|}
        \hline
        Effectiveness           & $F_1$-score   \\ \hline
        \texttt{RL4OASD}           &\textbf{0.854}        \\ \hline
        w/o noisy labels & 0.626         \\ \hline
        w/o road segment embeddings &  0.828       \\ \hline
        w/o RNEL  & 0.816    \\ \hline
        w/o DL  & 0.737    \\ \hline
        w/o local reward  &0.850         \\ \hline
        w/o global reward  &0.849         \\ \hline
        {\Revision{w/o ASDNet}} & {\Revision{0.508}}         \\ \hline
        {\Revision{only transition frequency}}  & {\Revision{0.643}}        \\ \hline
        \end{tabular}
\label{tab:ablation}
\vspace{-5mm}
\end{table}

\smallskip
\noindent \textbf{Evaluation Metrics.}
To verify the results of subtrajectory detection, we consider two evaluation metrics. \underline{First}, we adapt the evaluation metric $F_1$-score proposed for Named Entity Recognition (NER)~\cite{li2021modularized,li2021effective}. 
This is inspired by the fact that our task corresponds to one of tagging subsequences of a sequence, which is similar to NER, which tags phrases (i.e., subsequences) of a sentence (sequence).
The intuition is that we take the anomalous subtrajectories as entities in NER task. Specifically, (1) let $C_{g,i}$ denote a manually labeled anomalous subtrajectory $i$ in the ground truth set $C_g$, and $C_{o,i}$ denotes the corresponding subtrajectory $i$ in the set $C_o$, which is returned by a detection method. We employ Jaccard to measure the ratio of intersection over union of the road segments between $C_{g,i}$ and $C_{o,i}$. 
(2) We then measure the similarity between $C_{g}$ and $C_{o}$ by aggregating the Jaccard $\mathcal{J}_i(C_{g,i},C_{o,i})$.
\begin{align}
\mathcal{J}_i(C_{g,i},C_{o,i}) = \frac{|C_{g,i} \cap C_{o,i}|}{|C_{g,i} \cup C_{o,i}|}, \ \mathcal{J}(C_g, C_o) = \sum_{i=1}^{|C_g|} \mathcal{J}_i(C_{g,i},C_{o,i}).
\label{eq:jac}
\end{align}
Note that the intersection and union operations between $C_{g,i}$ and $C_{o,i}$ are based on 1's of the road segments of the manually labeled anomalous trajectory $i$. (3) Finally, we define the precision (P) and recall (R) by following~\cite{li2021modularized,li2021effective}, and compute the $F_1$-score accordingly.
\begin{align}
\text{P} = \frac{\mathcal{J}(C_g, C_o)}{|C_o|}, \quad \text{R} = \frac{\mathcal{J}(C_g, C_o)}{|C_g|}, \quad F_{1} = 2\times\frac{\text{P}\times\text{R}}{\text{P}+\text{R}}.
\label{eq:f1}
\end{align}
{\ICDERevision{\underline{Second}, we further design a variant of $F_1$-score. {\CHENG Specifically, we re-define the Jaccard similarity $\mathcal{J}_i(C_{g,i},C_{o,i})$ to be 1 if it is above a threshold $\phi$ and 0 otherwise.
Then, we compute the $F_1$-score by Equation~\ref{eq:f1} based on the re-defined Jaccard similarity. 
We call this variant of $F_1$-score $TF_1$-score, where the threshold $\phi$ is naturally set to 0.5 in this paper.
The intuition of $TF_1$-score is to count only those detected anomalous subtrajectories which are aligned with real anomalous subtrajectories sufficiently.}
}}

\smallskip
\noindent \textbf{Evaluation Platform.}
We implement \texttt{RL4OASD} and other baselines in Python 3.6 and Tensorflow 1.8.0. 
The experiments are conducted on a server with 10-cores of Intel(R) Core(TM) i9-9820X CPU @ 3.30GHz 64.0GB RAM and one Nvidia GeForce RTX 2080 GPU. The labeled datasets and codes can be downloaded via the link~\footnote{https://github.com/lizzyhku/OASD} to reproduce our work.

\subsection{Effectiveness Evaluation}
\label{sec:effectiveness}

\noindent \textbf{Comparison with existing baselines.} We study the anomalous subtrajectory detection with our labeled datasets. In Table~\ref{tab:baselines}, we report the effectiveness in terms of different trajectory lengths, e.g., we manually partition the Chengdu dataset into four groups, i.e., G1, G2, G3 and G4, such that the lengths in a groups are $G1<15$, $15 \leq G2 < 30$, $30 \leq G3 <45$ and $G4 \geq 45$. We also report the overall effectiveness in whole datasets. The results clearly show that \texttt{RL4OASD} consistently outperforms baselines in terms of different settings. {\ICDERevision{Specifically, it outperforms the best baseline (i.e., CTSS) for around 20\% and 15\%  (resp. 30\% and 28\%) in terms of $F_1$-score and $TF_1$-score in Chengdu (resp. Xi'an) regarding the overall effectiveness, and the improvement is around 5\%-26\% and 1\%-19\% (resp. 26\%-47\% and 22\%-45\%) in terms of $F_1$-score and $TF_1$-score in Chengdu (resp. Xi'an) for different groups. A possible reason is that 
CTSS needs a threshold to extract those anomalous parts with the anomaly scores larger than the threshold. However, the threshold is hard to set appropriately for all complex traffic cases. \texttt{RL4OASD} demonstrates its superioriority, which is mainly due to its data-driven nature for the anomalous subtrajectory detection task. Besides, we observe that \texttt{RL4OASD} performs similar trends of results in terms of $F_1$-score and $TF_1$-score, which shows the genericity of our method.
}}

\smallskip
\noindent \textbf{Ablation study.} We conduct an ablation study to show the effects of some components in \texttt{RL4OASD}, including (1) the noisy labels, (2) pre-trained road segment embeddings, (3) Road Network Enhanced Labeling (RNEL), (4) Delayed Labeling (DL), (5) local reward, (6) global reward, {\Revision{(7) ASDNet and (8) transition frequency.}}
In particular, for (1), we replace the noisy labels with random labels;
for (2), we randomly initialize the embeddings of road segments to replace the pre-trained embeddings provided by Toast~\cite{chen2021robust}; for (3), we drop the RNEL, and the model needs to take the action at each road segment without being guided by road networks; for (4), we drop the DL, and no delay mechanism is involved for labeling road segments; for (5), we drop the local reward, which encourages the continuity of refined labels; for (6), we drop the global reward, which provides some feedback indicating the quality of refined labels; {\Revision{for (7), we replace ASDNet with an ordinary classifier to follow the outputs of RSRNet, and train the classifier with noisy labels; for (8), we only use the transition frequency to detect anomalous subtrajectories, which can be regarded as the simplest method.}}

{\addComment{In Table~\ref{tab:ablation}, we observe each component benefits the overall effectiveness, where the ASDNet contributes quite much, because it is utilized to refine the labels which are utilized to train the RSRNet, and find out the anomalous subtrajectories. Besides, we note that (1) noisy labels, (4) Delayed Labeling (DL) and (8) transition frequency also contribute much to the performance of \texttt{RL4OASD}. As the aforementioned, noisy labels provide a necessary warm-start before the training with the improvement of around 36\%, the delayed labeling provides the continuity of anomalous subtrajectories with the improvement of around 16\%, and if we only use the transition frequency for the detection task, the performance degrades a lot by around 25\%. In addition, we also verify the effect of (2) pre-trained road segment embeddings, (3) RNEL, (5) local reward and (6) global reward. From Table ~\ref{tab:ablation}, we observe that road segment embeddings benefit the overall effectiveness by providing the traffic context from road networks, and local reward and global reward provide some feedback signals, which guide the training of RSRNet to further provide better states for ASDNet. In addition, with the RNEL, we notice an effectiveness improvement of around 5\%, since it simplifies the cases of making decisions and the model becomes easier to train. 
}}

\subsection{Parameter Study}

\noindent \textbf{Varying parameter $\alpha$, $\delta$ and $D$.} We study the effects of parameter $\alpha$ for constructing noisy labels, parameter $\delta$ for constructing normal route features and parameter $D$ for controlling the number of delayed road segments in the delaying mechanism. The results and detailed description are put into the technical report~\cite{TR} due to the page limit. Overall, we observe that a moderate setting {\CHENG (with $\alpha = 0.5$, $\delta = 0.4$, and $D = 8$)} contributes to the best effectiveness.

\begin{figure}[t!]
\centering
\begin{tabular}{c}
  \begin{minipage}{0.93\linewidth}
    \includegraphics[width=1.02\linewidth]{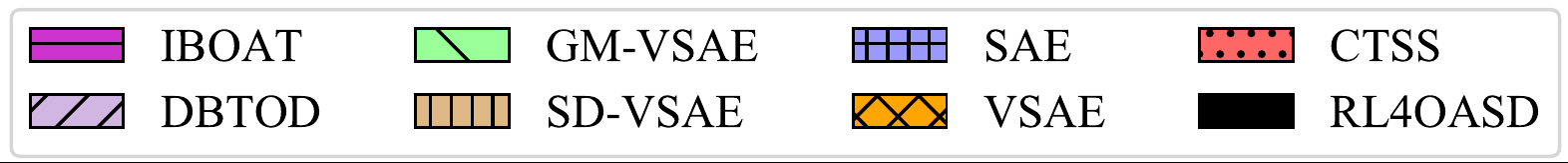}
   \end{minipage}
   \\
  \begin{minipage}{0.93\linewidth}
    \includegraphics[width=1.02\linewidth]{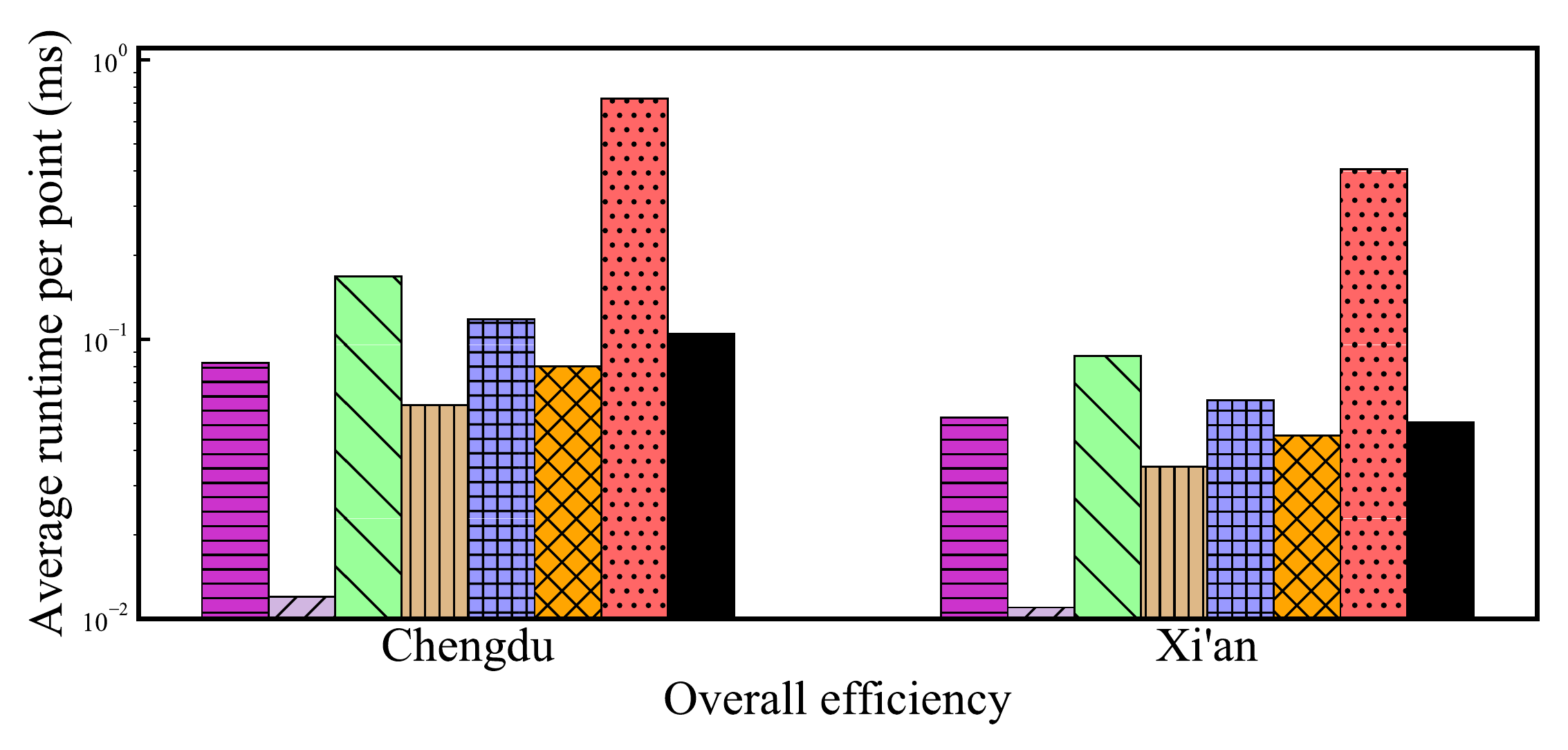}
  \end{minipage}
\end{tabular}
\vspace*{-2mm}
\caption{Overall detection efficiency.}
\label{fig:overall_efficiency}
\vspace*{-5mm}
\end{figure}

\subsection{Efficiency Study}
\noindent\textbf{Overall detection.} Figure~\ref{fig:overall_efficiency} reports the online detection efficiency in terms of average running time per point on both Chengdu and Xi'an datasets. We provide the detailed analysis as follows. 

We observe the running time in Chengdu is larger than in Xi'an, because the trajectories are generally shorter in Xi'an. {\Revision{We observe DBTOD runs the fastest on both datasets, because it is a light model with low-dimensional embeddings of some cheaper features such as road-level and turning angle for the detection, which can be accomplished very quickly, while \texttt{RL4OASD} involves more operations, including an LSTM-based RSRNet to capture features, and an RL-based ASDNet to label each road segment.}} CTSS runs the slowest since it involves discrete Frechet distance to compute the deviation between a target trajectory and a given reference trajectory, which suffers from a quadratic time complexity. {\Revision{In addition, for four learning-based methods GM-VSAE, SD-VSAE, SAE and VSAE that are proposed for trajectory detection via the generation scheme, we observe SD-VSAE and VSAE are generally faster than the others (i.e., GM-VSAE and SAE). This is because SAE is proposed with a traditional Seq2Seq structure, where it involves the operations of encoding and decoding, which needs to scan a trajectory twice. Compared with SAE, VSAE is free of the encoding step, and only involves the decoding step to detect some possible anomalies. For SD-VSAE, it is a fast version of GM-VSAE, where it only predicts one Gaussian component in the encoding (or inference) step with its SD module; however, GM-VSAE needs several components in the encoding step, and uses all of them to detect anomalies in the decoding. The results are consistent with the findings that are reported in~\cite{liu2020online}.}} Overall, \texttt{RL4OASD} runs reasonably fast and would meet the practical needs, e.g., it takes less than 0.1ms to process each point, which is 20,000 times faster than the practical sampling rate of the trajectory data (2s).

\begin{figure}[h]
	\vspace{-3mm}
	\centering
	\begin{tabular}{c c}
		\begin{minipage}{3.7cm}
			\includegraphics[width=4cm]{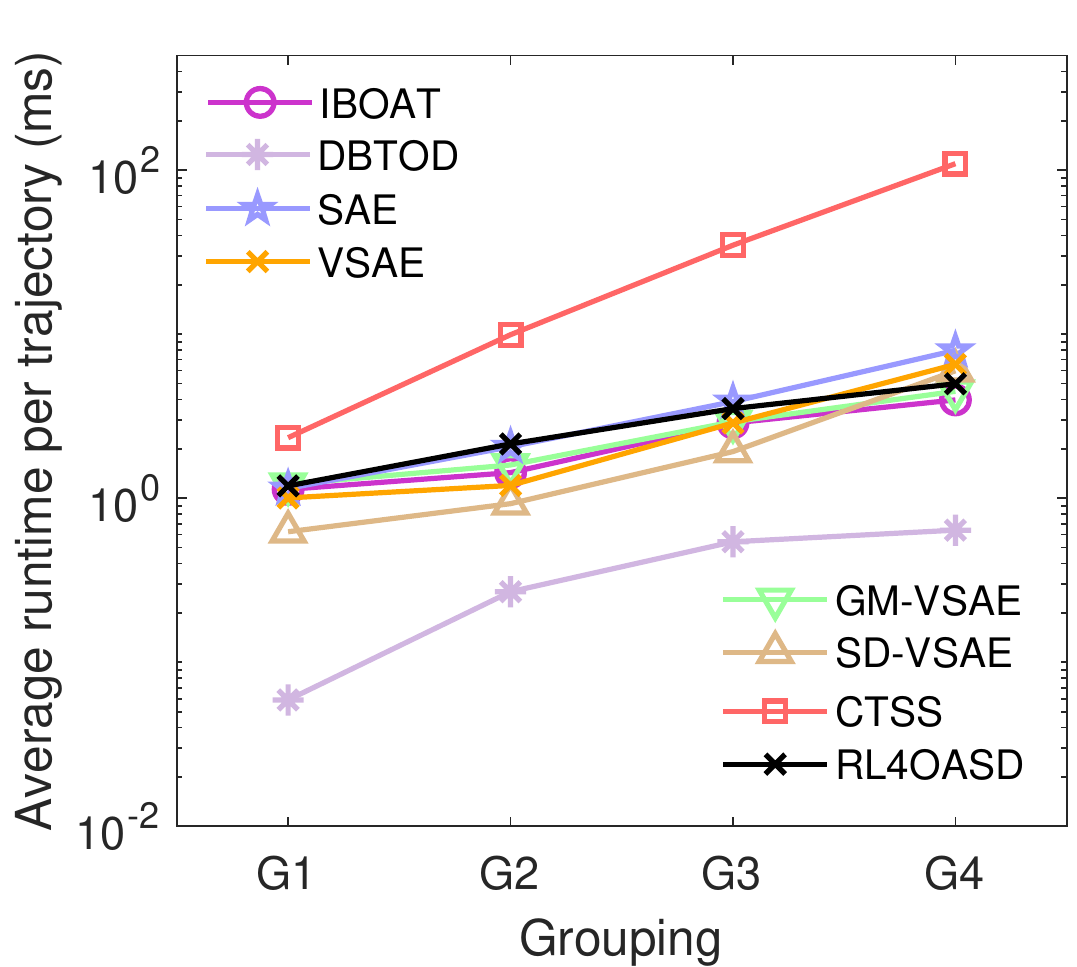}
		\end{minipage}
		&
		\begin{minipage}{3.7cm}
			\includegraphics[width=4cm]{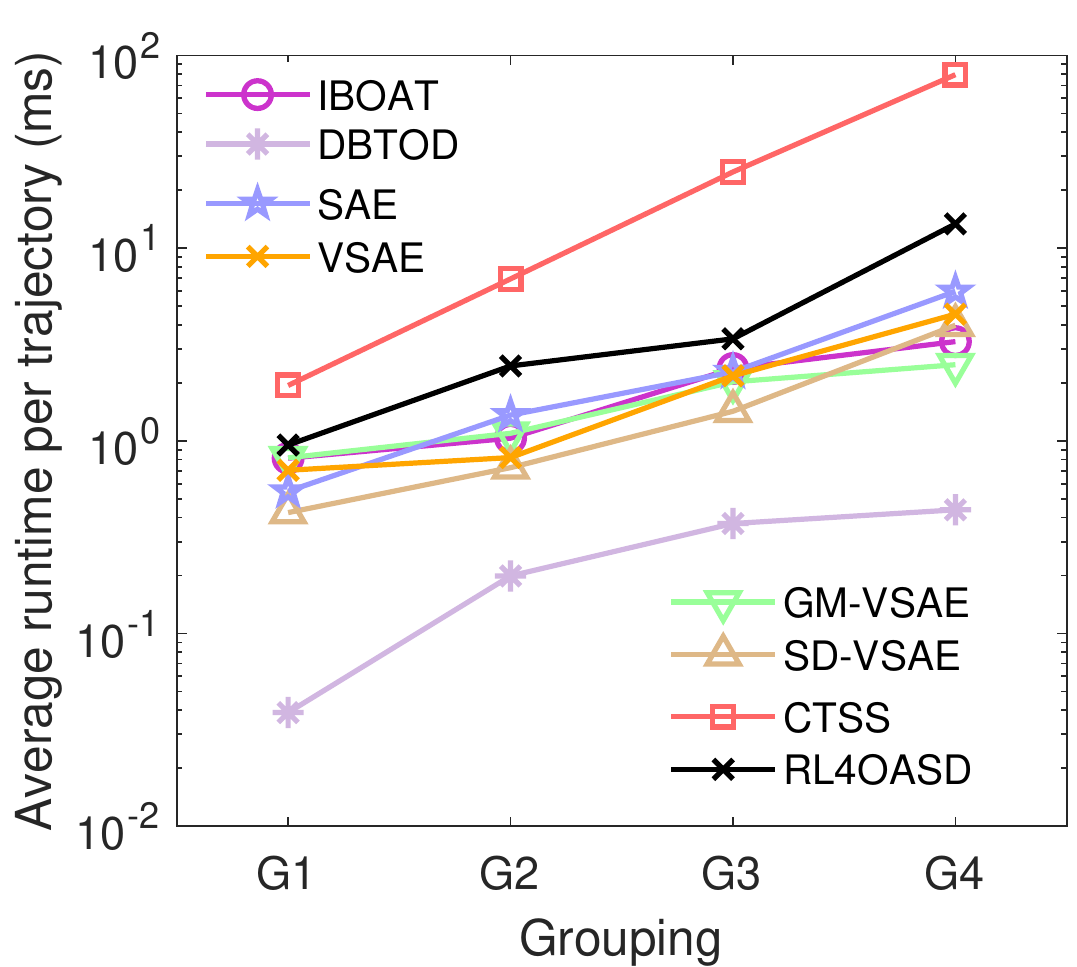}
		\end{minipage}
		\\
		(a) Chengdu
		&
		(b) Xi'an
	\end{tabular}
	\vspace*{-2mm}
	\caption{Detection scalability.}\label{fig:length}
	\vspace{-4mm}
\end{figure}

\smallskip
\noindent \textbf{Detection scalability.} Further, we study the scalability for detecting ongoing trajectories in terms of four trajectory groups with different lengths. In Figure~\ref{fig:length}, we report the average running time per trajectory on both datasets. {\Revision{We observe the CTSS runs slowest, and its disparity between others becomes larger as the trajectory length increases. This is because CTSS involves the trajectory similarity with discrete Frechet distance to detect possible anomalies, the trend is consistent with its time complexity. DBTOD is a light model, which consistently runs the fastest. However, in terms of the effectiveness comparison reported in Table~\ref{tab:baselines}, we note that DBTOD is not a good model for the subtrajectory detection task, though it runs faster than \texttt{RL4OASD}. In general, \texttt{RL4OASD} shows similar trends as others, and scales well with the trajectory length grows. This is because \texttt{RL4OASD} involves the graph structure of road networks for labeling road segments (i.e., RNEL), and the time of taking actions can be saved accordingly.}}

\begin{figure}
\centering
\scriptsize
\hspace{-4mm}
\begin{tabular}{c c c}
  \begin{minipage}{0.34\linewidth}
  \includegraphics[width=\linewidth]{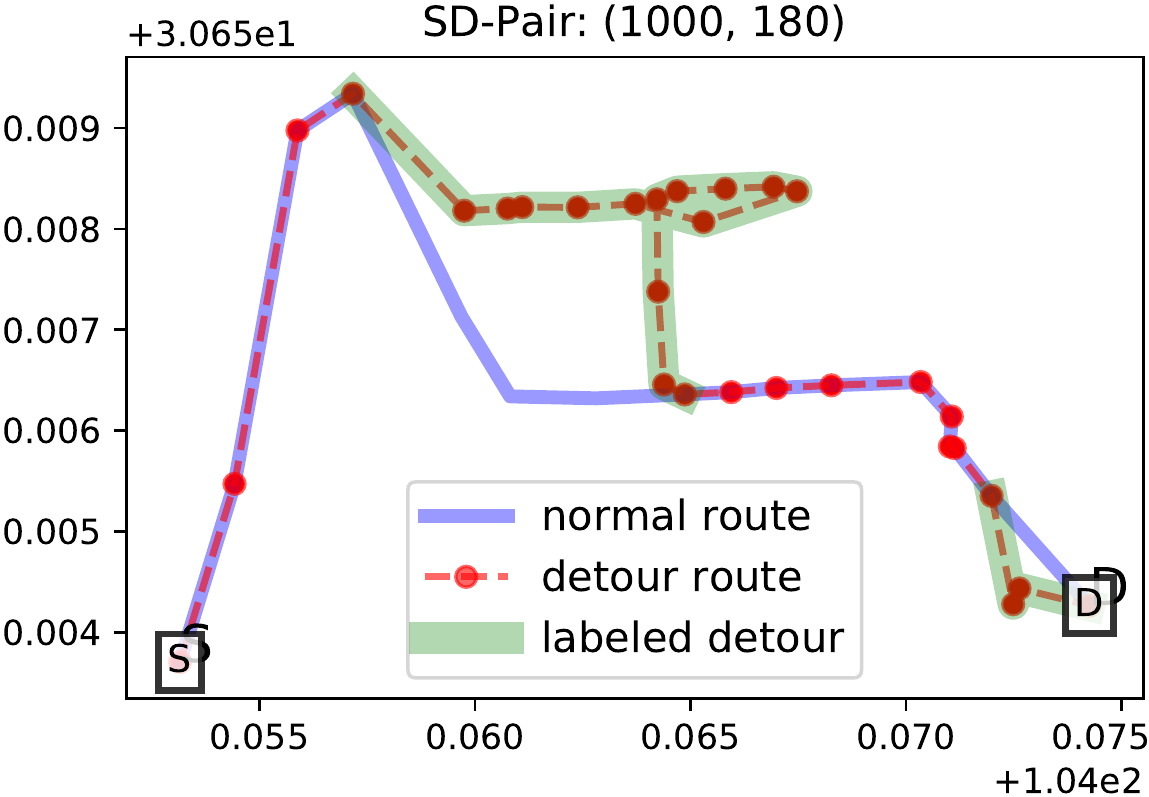}
  \end{minipage}\hspace{-4mm}
  &
  \begin{minipage}{0.34\linewidth}
    \includegraphics[width=\linewidth]{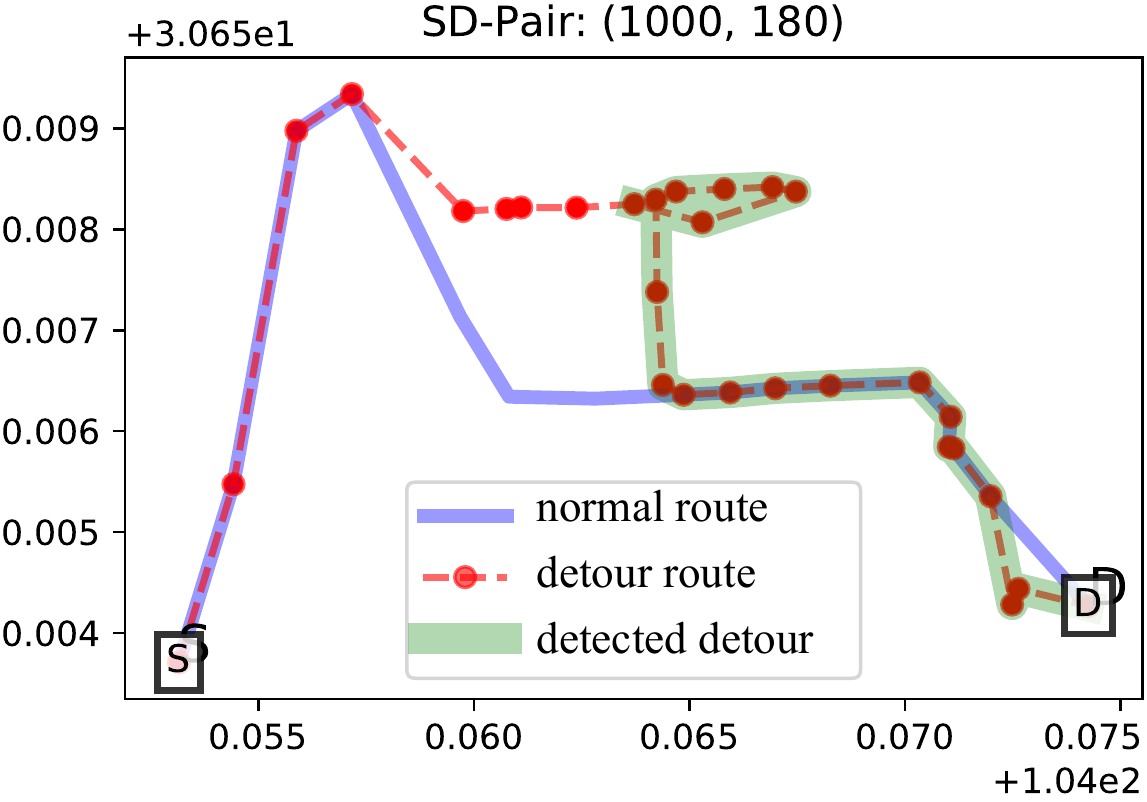}
  \end{minipage}\hspace{-4mm}
  &
  \begin{minipage}{0.34\linewidth}
  \includegraphics[width=\linewidth]{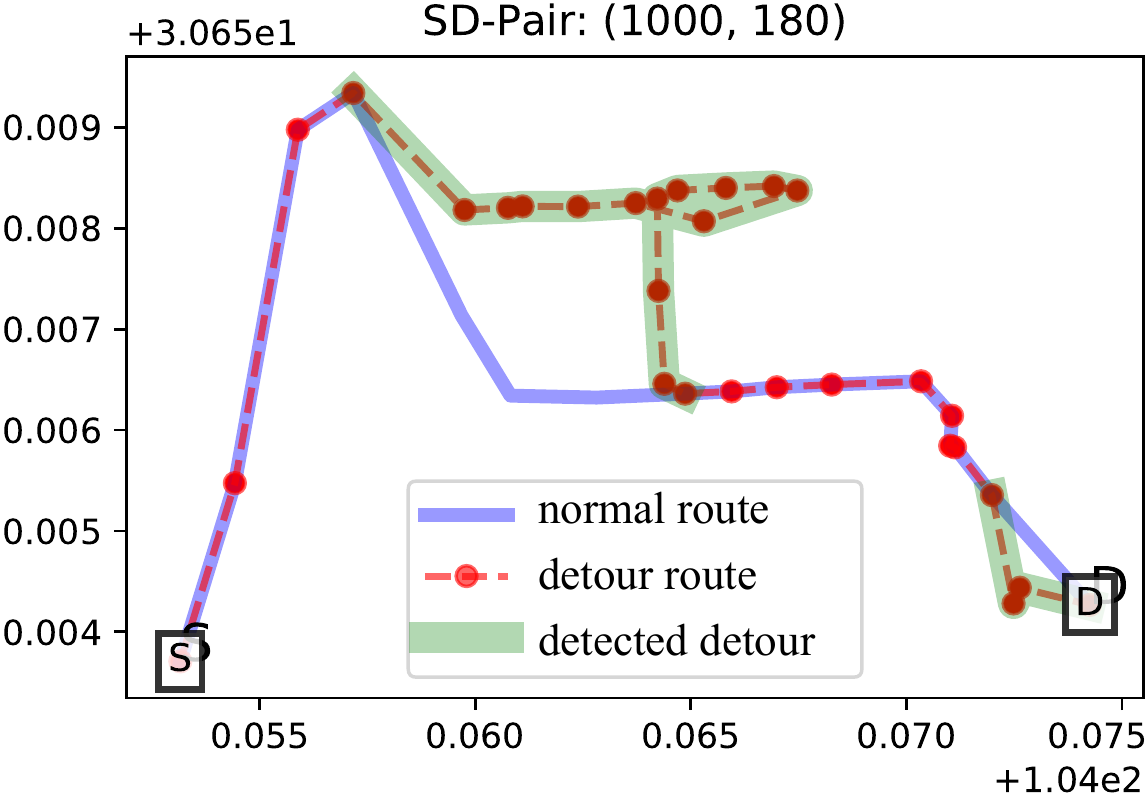}%
  \end{minipage}\hspace{-4mm}
  \\
  (a) Ground truth ($F_1$=1.0)
   &
  (b) CTSS ($F_1$=0.792)
   &
  (c) \texttt{RL4OASD} ($F_1$=1.0)
\end{tabular}
\caption{Case study, blue lines: the normal routes traveled by most of trajectories; red dashed lines: the routes contained anomalous subtrajectories; red points: the intersections on road networks; green lines: the detected anomalous subtrajectories. 
}
\label{fig:casestudy}
\vspace*{-2mm}
\end{figure}

\subsection{Case Study}
We investigate representative detour cases in Chengdu. We visualize with the green lines the detours labeled as/by Ground truth, CTSS and \texttt{RL4OASD}. Here, we choose CTSS for comparison, since it shows the best effectiveness among baselines. In Figure~\ref{fig:casestudy}, it illustrates the case of two detours in a route, we observe \texttt{RL4OASD} detects the detours accurately with $F_1$-score=1.0; however, CTSS fails to detect the starting position of the detection. This is because CTSS measures the trajectory similarity via Frechet distance between the normal trajectory (blue) and the current ongoing trajectory (red) at each timestamp, where a detour may already happen though the ongoing trajectory is still similar to the reference at several starting positions. More results are illustrated in~\cite{TR}.

\begin{table}[]
\setlength{\tabcolsep}{5pt}
\vspace{-3mm}
{\ICDERevision{\caption{Preprocessing and training time, the map matching~\cite{yang2018fast} is in C++, noisy labeling and training are in Python.
}
\label{tab:train_time}}}
\vspace{-3mm}
\begin{tabular}{|cc|c|c|c|c|c|}
\hline
\multicolumn{2}{|c|}{Data size}                                                                                                                                       & 4,000 & 6,000 & 8,000 & 10,000         & 12,000         \\ \hline
\multicolumn{1}{|c|}{\multirow{2}{*}{\begin{tabular}[c]{@{}c@{}}{\ICDERevision{Preprocessing}} \\ {\ICDERevision{time}}\end{tabular}}} & \begin{tabular}[c]{@{}c@{}}{\ICDERevision{Map}} \\ {\ICDERevision{matching (s)}}\end{tabular}   & {\ICDERevision{15.82}} & {\ICDERevision{23.17}} &{\ICDERevision{30.41}} &{\ICDERevision{39.95}}          &{\ICDERevision{48.74}}          \\ \cline{2-7} 
\multicolumn{1}{|c|}{}                                                                                & \begin{tabular}[c]{@{}c@{}}{\ICDERevision{Noisy}} \\ {\ICDERevision{labeling (s)}}\end{tabular} & {\ICDERevision{20.31}} & {\ICDERevision{28.40}} & {\ICDERevision{36.25}} & {\ICDERevision{41.21}}          & {\ICDERevision{46.82}}          \\ \hline
\multicolumn{2}{|c|}{Training time (hours)}                                                                                                                           & 0.10  & 0.14  & 0.17  & 0.21           & 0.25           \\ \hline
\multicolumn{2}{|c|}{$F_1$-score}                                                                                                                                     & 0.723 & 0.786 & 0.821 & \textbf{0.854} & \textbf{0.854} \\ \hline
\end{tabular}
\vspace*{-3mm}
\end{table}

{\ICDERevision{\subsection{Preprocessing and Training Time}
With the default setup in Section~\ref{sec:setup}, we study the preprocessing time (i.e., map-matching and noisy labeling) and training time of \texttt{RL4OASD} in Table~\ref{tab:train_time}, by varying the data size from 4,000 trajectories to 12,000 trajectories, and report the time costs and $F_1$-scores. 
Overall, it takes less than two minutes in the preprocessing and around 0.2 hours to obtain a satisfactory model, and the scale is linear with the data size, which indicates the capability of \texttt{RL4OASD} for large-scale trajectory data. Here, we choose 10,000 trajectories to train \texttt{RL4OASD}, which provides a reasonable trade-off between effectiveness and training time cost.
}}

\begin{figure}[h]
	\hspace{-.8cm}
	\small
	\centering
	\begin{tabular}{c c}
		\begin{minipage}{3.8cm}
			\includegraphics[width=3.9cm]{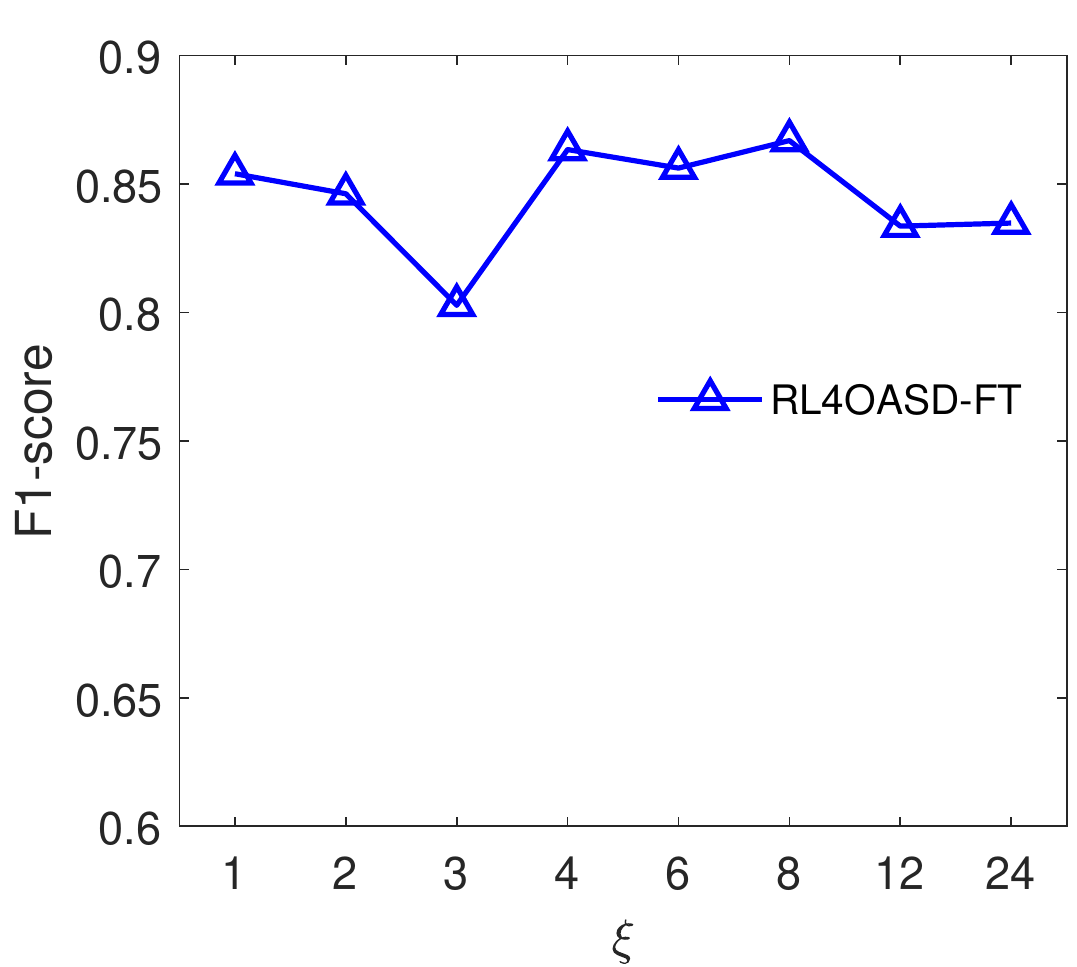}
		\end{minipage}
		&
		\begin{minipage}{3.8cm}
			\includegraphics[width=3.9cm]{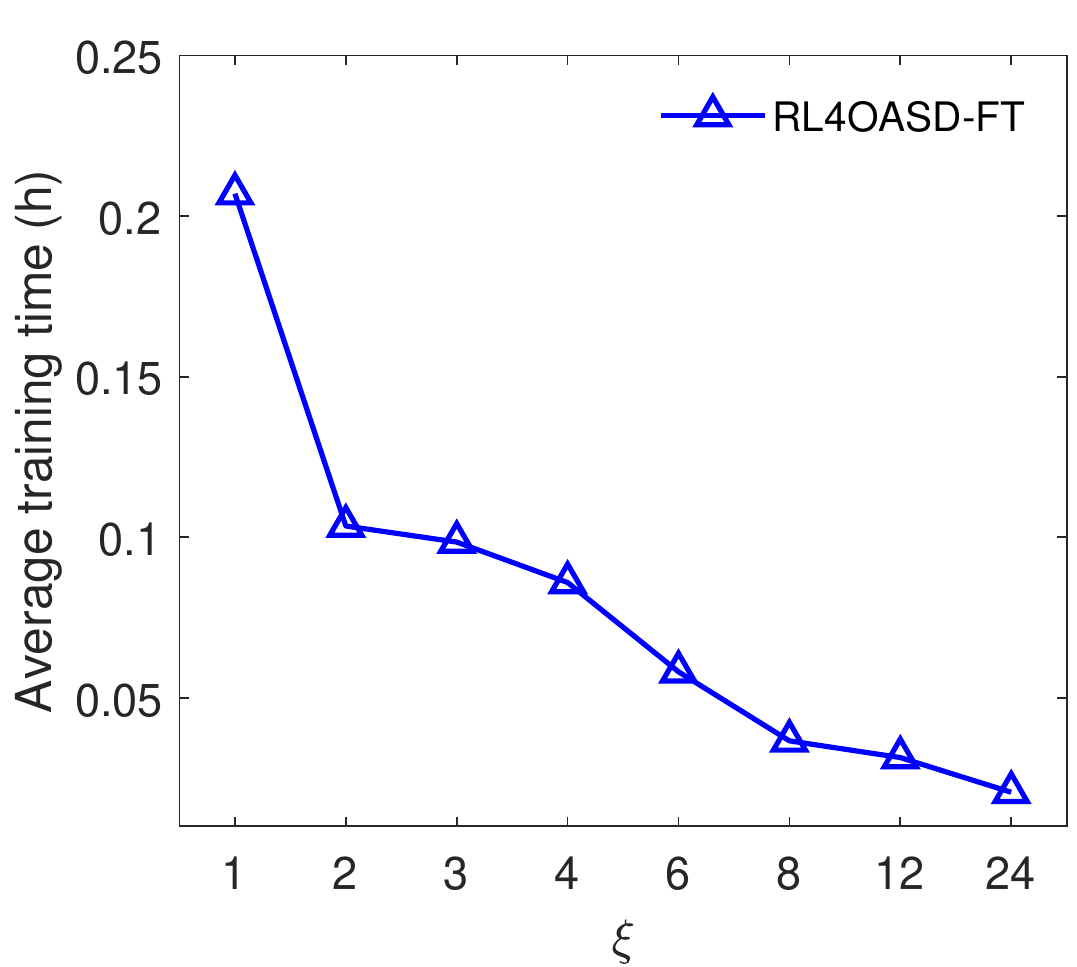}
		\end{minipage}
		\\
		(a) $F_1$-score varying $\xi$
		&
		(b) Training time varying $\xi$
		\\
		\begin{minipage}{3.8cm}
			\includegraphics[width=3.9cm]{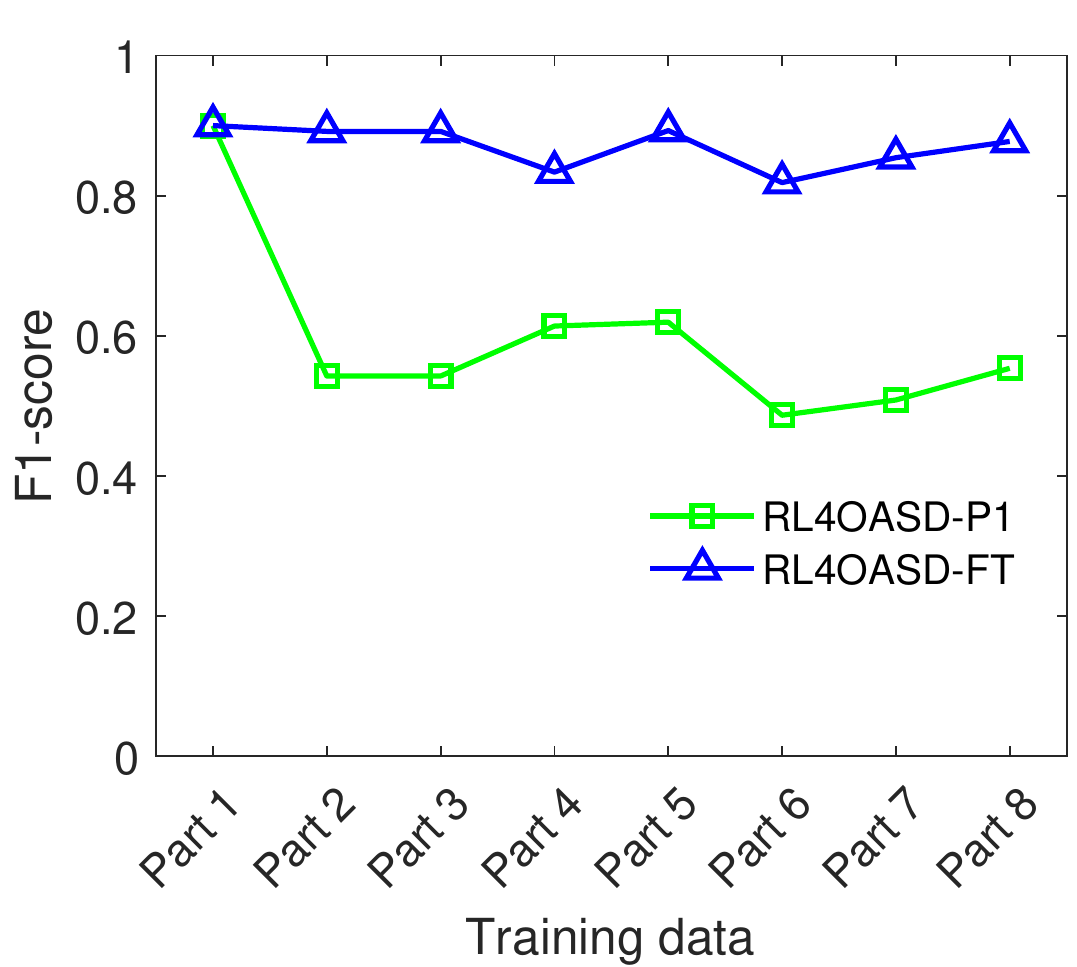}
		\end{minipage}
		&
		\begin{minipage}{3.8cm}
			\includegraphics[width=3.9cm]{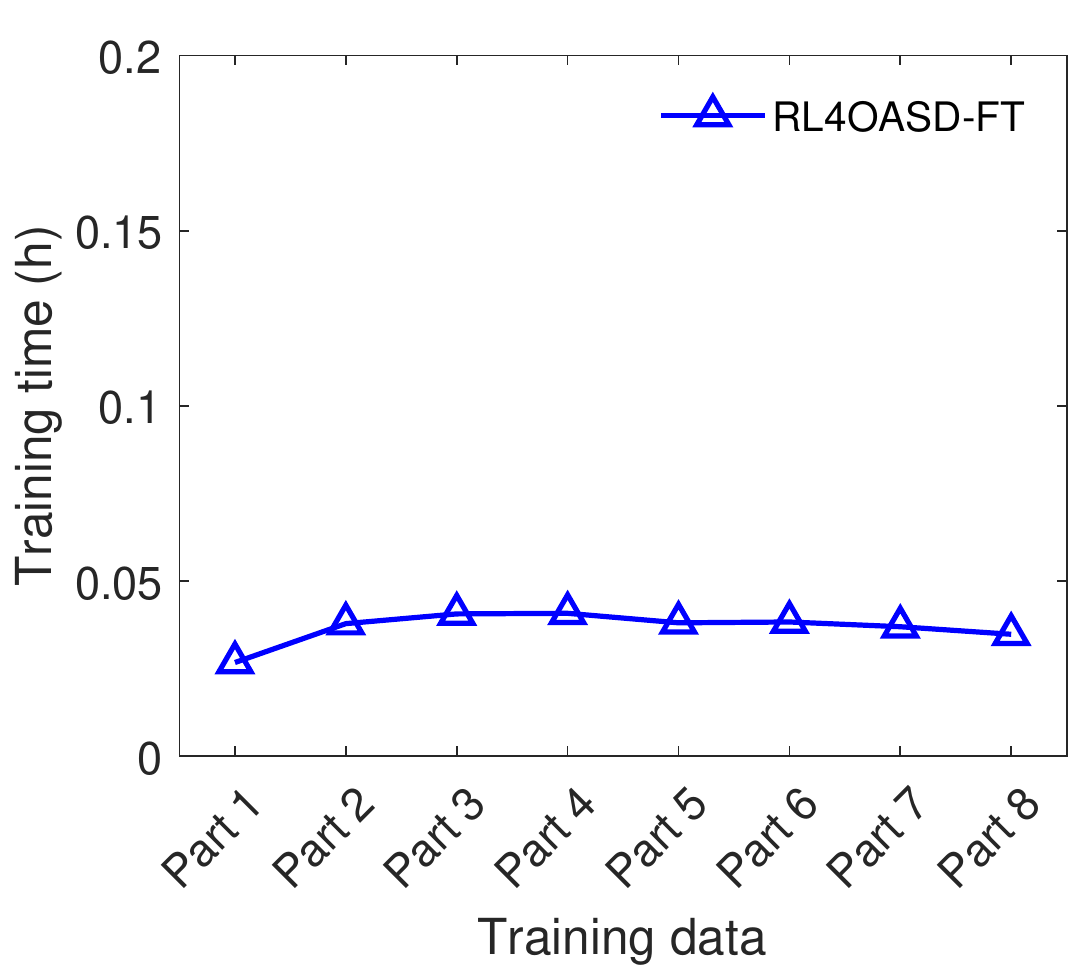}
		\end{minipage}
		\\
		(c) $F_1$-score ($\xi=8$)
		&
		(d) Training time ($\xi=8$)
	\end{tabular}
	\vspace*{-1mm}
	{\ICDERevision{\caption{{\Revision{\texttt{RL4OASD} with or without fine-tuning and the training times.}}}\label{fig:traffic}}}
	\vspace{-3mm}
\end{figure}

{\ICDERevision{\subsection{Detection in Varying Traffic Conditions}
\label{sec:concept_exp}
By following the setting in~\cite{liu2020online}, we first sort the trajectories by their starting timestamps within one day, and split all the sorted trajectories into $\xi$ partitions. 
{\CHENG For example, when $\xi$ is set 4, we would have 4 parts within one day from Part 1 (i.e., 00:00 - 06:00, the earliest) to Part 4 (i.e., 18:00 - 24:00, the latest).}
Here, we consider two models, namely \texttt{RL4OASD-P1} and \texttt{RL4OASD-FT}, to show the effectiveness of our online learning strategy for the varying traffic conditions. For \texttt{RL4OASD-P1}, we train it on Part 1 and directly apply it to all parts. For \texttt{RL4OASD-FT}, we train it on Part 1, and keep updating it (i.e., fine-tuning using stochastic gradient descent) for other parts. 

We first study how to set a proper $\xi$.
We vary the $\xi$ from 1 to 24, and the average $F_1$-scores over all parts and the corresponding training times are reported in Figure~\ref{fig:traffic}(a) and (b), respectively. We observe that the performance fluctuates as $\xi$ increases, and the corresponding training time decreases because with more data parts, the size of training data for each part becomes less. 
We choose the $\xi=8$, since it leads to the best effectiveness (i.e., $F_1$-score=0.867) with the average training time below 0.04 hours, which is much smaller than the duration of a time period (i.e., $24/8=3$ hours).

With the setting of $\xi=8$, we further compare \texttt{RL4OASD-FT} with \texttt{RL4OASD-P1}, and report the $F_1$-score for each part in Figure~\ref{fig:traffic}(c). We observe that the performance of \texttt{RL4OASD-P1} degrades on Part 2–7, which can be attributed to the concept drift. In contrast, \texttt{RL4OASD-FT} improves the performance when new trajectories are being recorded to train. 
In addition, our model is also able to handle situations with the traffic condition updated very frequently (e.g., hourly) by training from the newly recorded trajectory data. To see this, we report the training time for each part in Figure~\ref{fig:traffic}(d). It normally takes below 0.05 hours to update the model for each part, which largely meets the practical needs (e.g., far below the duration of each time period).}}

\begin{figure}
\centering
\begin{tabular}{c c}
  \begin{minipage}{0.45\linewidth}
  \includegraphics[width=\linewidth]{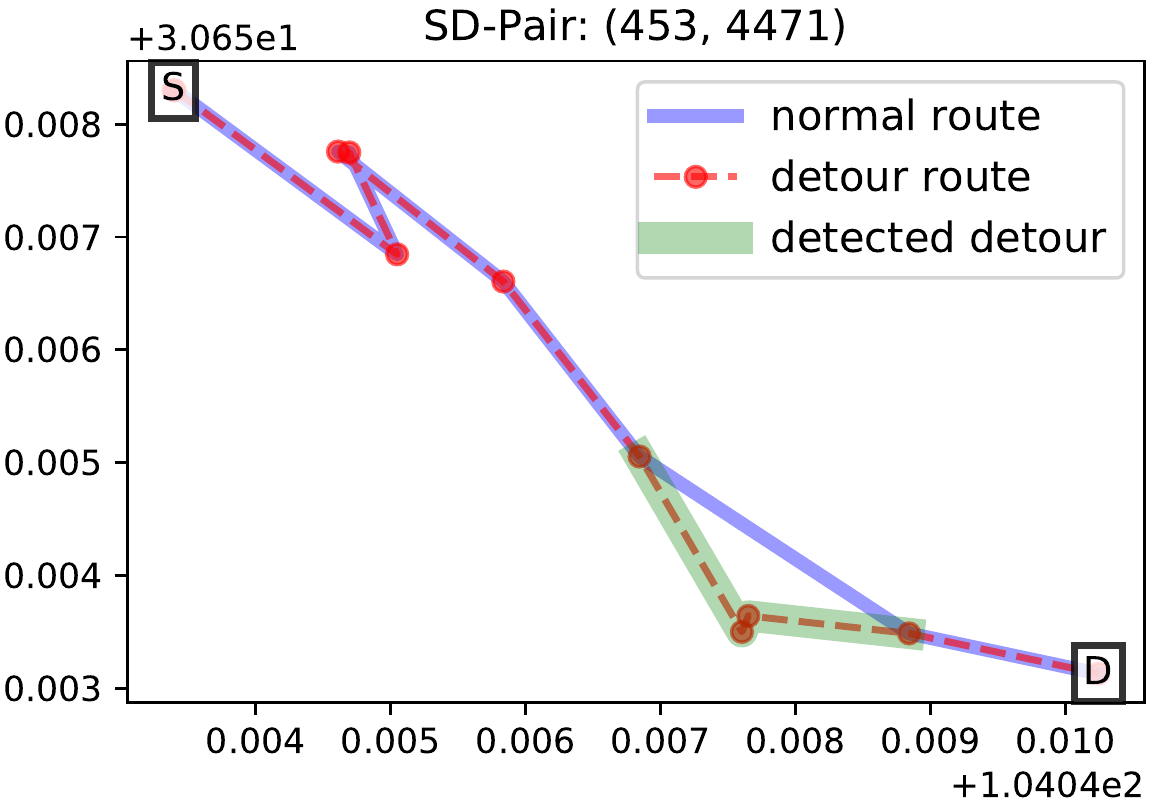}%
  \end{minipage}
  &
  \begin{minipage}{0.45\linewidth}
  \includegraphics[width=\linewidth]{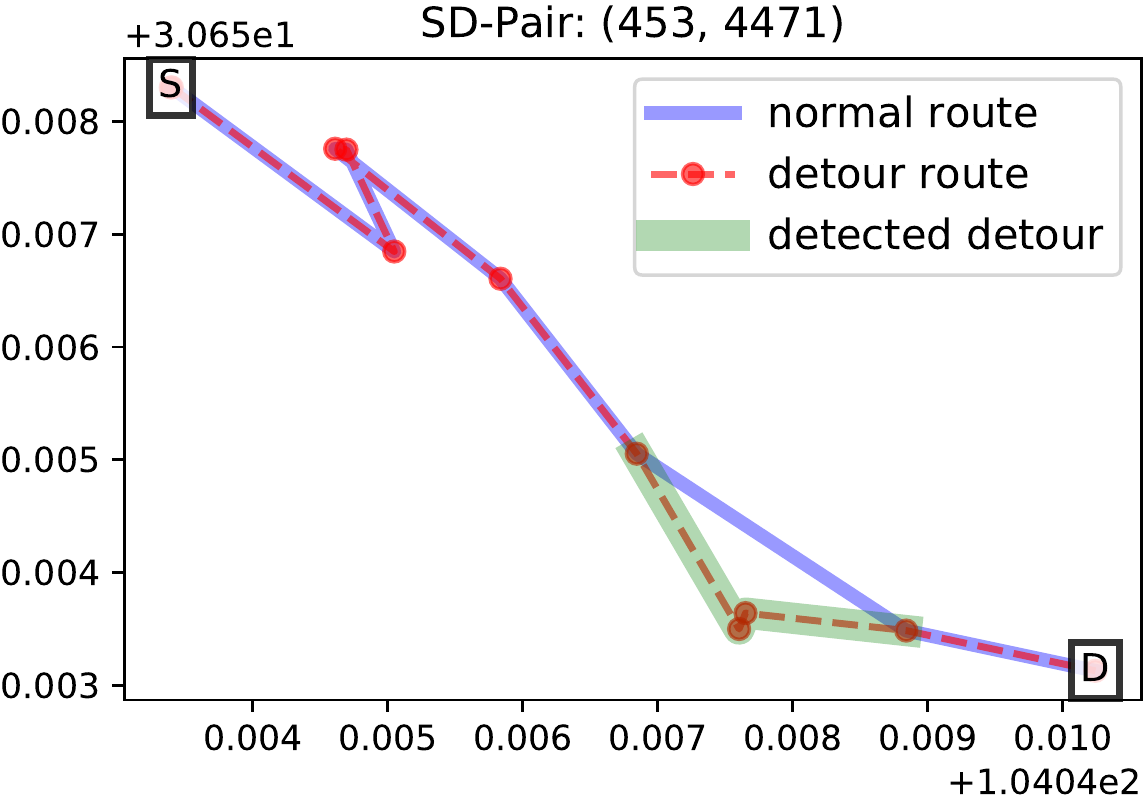}%
  \end{minipage}
  \\
  (a) \texttt{RL4OASD-P1}, $F_1$=1.0
  &
  (b) \texttt{RL4OASD-FT}, $F_1$=1.0
  \\
 \begin{minipage}{0.45\linewidth}
 \includegraphics[width=\linewidth]{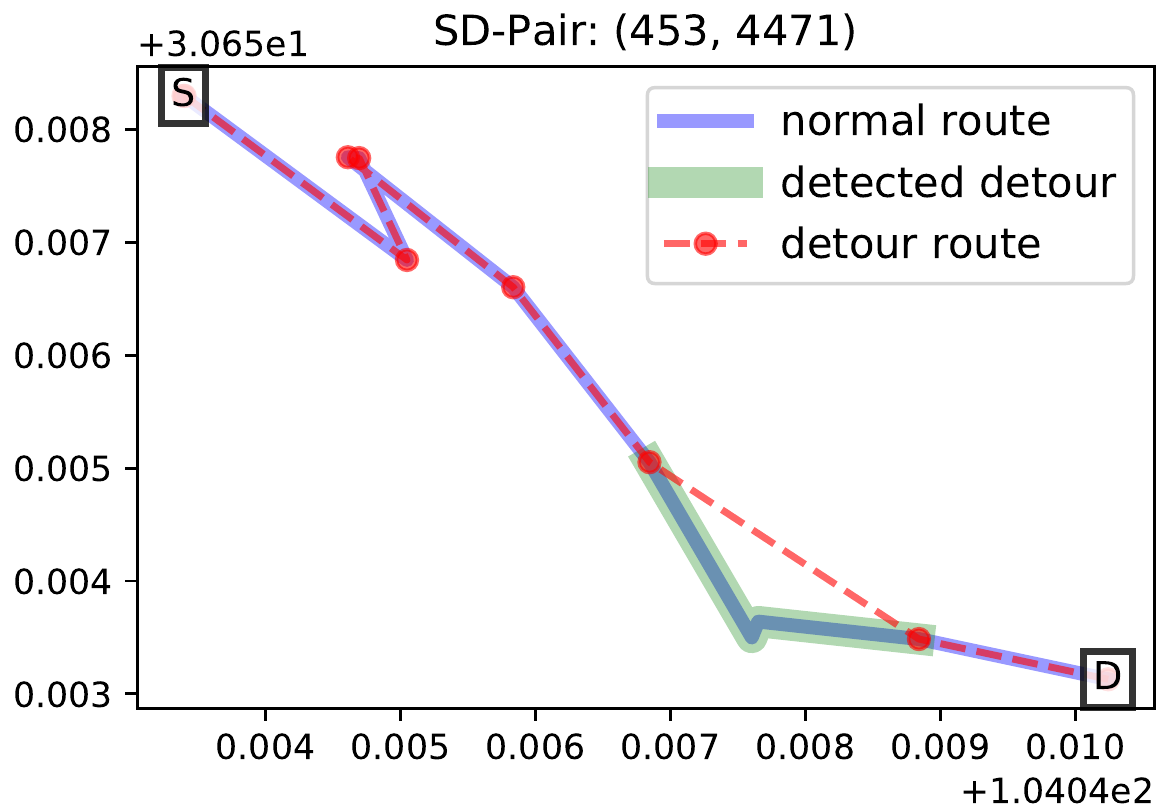}
 \end{minipage}
 &
 \begin{minipage}{0.45\linewidth}
 \includegraphics[width=\linewidth]{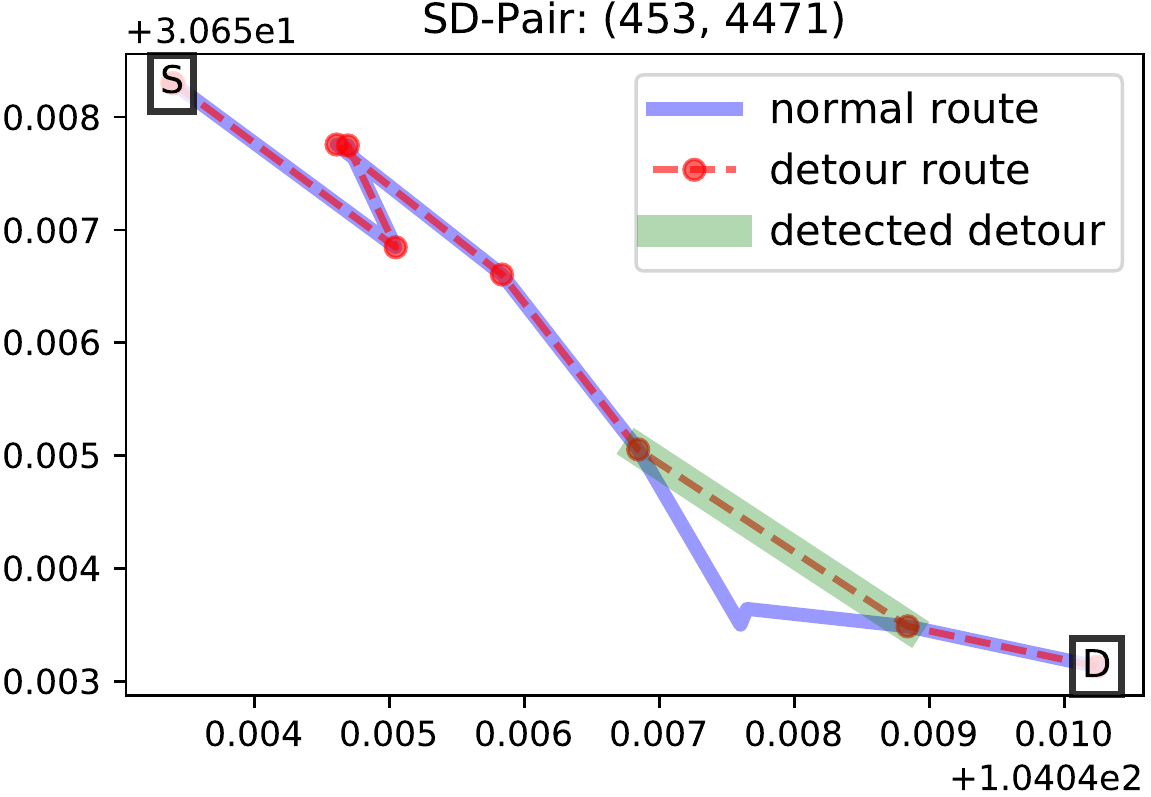}%
 \end{minipage}
 \\
  (c) \texttt{RL4OASD-P1}, $F_1$=0.78
  &
  (d) \texttt{RL4OASD-FT}, $F_1$=1.0
\end{tabular}
\vspace*{-3mm}
\caption{Concept drift, where (a)-(b) for Part 1 and (c)-(d) for Part 2.}
\label{fig:casestudy_drift}
\vspace*{-5mm}
\end{figure}

\begin{table}[t]
\setlength{\tabcolsep}{10pt}
\centering
\vspace{-2mm}
{\ICDERevision{\caption{Cold-start problem in RL4OASD.}
\label{tab:droprate}}}
\vspace{-3mm}
{\ICDERevision{\begin{tabular}{|c|c|c|c|c|c|c|}
\hline
Drop rate & 0.0 & 0.2 & 0.4 & 0.6 & 0.8 \\ \hline
$F_1$-score &\textbf{0.854} &\textbf{0.854} &0.852 &0.831 &0.803 \\ \hline
\end{tabular}}}
\vspace*{-4mm}
\end{table}

We also conduct a case study to illustrate the effectiveness of \texttt{RL4OASD-FT} in Figure~\ref{fig:casestudy_drift}. We observe the normal route and anomalous route are exchanged from Part 1 to Part 2. With the online learning strategy, \texttt{RL4OASD-FT} can still detect the detour on Part 2 with $F_1$-score=1.0; however, \texttt{RL4OASD-P1} causes the false-positive in Figure~\ref{fig:casestudy_drift}(c), because its policy is only trained on Part 1.

{\ICDERevision{\subsection{Cold-start Problem with Insufficient Historical Trajectories}
\label{sec:cold_start}
We study the effect of cold-start problem, where we vary the number of historical trajectories within the SD pairs with a drop rate parameter. For example, if the drop rate is set to 0.2, we randomly remove 20\% trajectories for the SD pairs, and the corresponding results are reported in Table~\ref{tab:droprate}. We observe that the effectiveness of \texttt{RL4OASD} is not affected by the cold-start problem much. For example, the model only degrades by 6\% when 80\% of historical trajectories are removed. This is possibly because the normal route feature is computed in a relative way (e.g., by a fraction between 0 and 1), {\CHENG and the normal routes can still be identified when only few trajectories within a SD pair are available}.
}}

\section{CONCLUSIONS}
\label{sec:conclusion}
In this paper, {\Revision{we study the problem of online anomalous subtrajectory detection on road networks and propose the first deep reinforcement learning based solution called \texttt{RL4OASD}. 
\texttt{RL4OASD} is a data-driven approach that can be trained without labeled data.
We conduct extensive experiments on two real-world taxi trajectory datasets with manually labeled anomalies. The results demonstrate that \texttt{RL4OASD} consistently outperforms existing algorithms, runs comparably fast, and 
supports the detection in varying traffic conditions.}} 
{\ICDERevision{In the future, we will explore the cold-start problem when the historical trajectories are not sufficient.
}}

\smallskip
\noindent\textbf{Acknowledgments:}
The project is partially supported by the funding from HKU-SCF FinTech Academy and the ITF project (ITP/173/18FP). This research/project is supported by the National Research Foundation, Singapore under its AI Singapore Programme (AISG Award No: AISG-PhD/2021-08-024[T] and AISG Award No: AISG2-TC-2021-001).
This research is also supported by the Ministry of Education, Singapore, under its Academic Research Fund (Tier 2 Awards MOE-T2EP20220-0011 and MOE-T2EP20221-0013). Any opinions, findings and conclusions or recommendations expressed in this material are those of the author(s) and do not reflect the views of National Research Foundation, Singapore and Ministry of Education, Singapore.




\bibliography{ref}
\bibliographystyle{IEEEtran}

\end{document}